\documentclass[aps,prd,superscriptaddress,twoside,twocolumn,nofootinbib,10pt,%
showpacs,floatfix]{revtex4-1}

\usepackage{amsmath,amssymb}
\usepackage{graphicx,bm}
\usepackage{dcolumn}
\usepackage{epstopdf}
\usepackage{ulem} 
\usepackage{subfigure}
\usepackage{float}
\usepackage[usenames]{color}

\allowdisplaybreaks

\begin{document}

\title{Dense baryonic matter in conformally-compensated hidden local symmetry: \\
Vector manifestation and chiral symmetry restoration}

\author{Yong-Liang Ma}
\email{yongliangma@jlu.edu.cn}
\affiliation{Department of Physics, Nagoya University, Nagoya, 464-8602, Japan}
\affiliation{College of Physics, Jilin University, Changchun, 130012, China}
\affiliation{Department of Physics, Kyungpook National University, Daegu 702-701, Korea}

\author{Masayasu Harada}
\email{harada@hken.phys.nagoya-u.ac.jp}
\affiliation{Department of Physics, Nagoya University, Nagoya, 464-8602, Japan}

\author{Hyun Kyu Lee}
\email{hyunkyu@hanyang.ac.kr}
\affiliation{Department of Physics, Hanyang University, Seoul 133-791, Korea}

\author{Yongseok Oh}
\email{yohphy@knu.ac.kr}
\affiliation{Department of Physics, Kyungpook National University, Daegu 702-701, Korea}
\affiliation{Asia Pacific Center for Theoretical Physics, Pohang, 
Gyeongbuk 790-784, Korea}

\author{Byung-Yoon Park}
\email{bypark@cnu.ac.kr}
\affiliation{Department of Physics, Chungnam National University, Daejeon 305-764, Korea}

\author{Mannque Rho}
\email{mannque.rho@cea.fr}
\affiliation{Department of Physics, Hanyang University, Seoul 133-791, Korea}
\affiliation{Institut de Physique Th\'eorique, CEA Saclay, 91191 Gif-sur-Yvette c\'edex, France}

\date{\today }
\begin{abstract}
We find that, when the dilaton is implemented as a (pseudo-)Nambu-Goldstone boson using
a conformal compensator or ``conformon" in a hidden gauge symmetric Lagrangian written to
$O(p^4)$ from which baryons arise as solitons, namely, skyrmions, the vector manifestation
and chiral symmetry restoration at high density predicted in hidden local symmetry theory
--- which is consistent with Brown-Rho scaling --- are lost or sent to infinite density.
It is shown that they can be restored if in medium the behavior of the $\omega$ field
is taken to deviate from that of the $\rho$ meson in such a way that the flavor $U(2)$
symmetry is strongly broken at increasing density.
The hitherto unexposed crucial role of the $\omega$ meson in the structure of elementary
baryon and multibaryon systems is uncovered in this work.
In the state of half-skyrmions to which the skyrmions transform at a density
$n_{1/2}^{} \gtrsim n_0^{}$ (where $n_0^{}$ is the normal nuclear matter density),
characterized by the vanishing (space averaged) quark condensate but nonzero pion decay
constant, the nucleon mass remains more or less constant at a value $\gtrsim$ 60 \% of
the vacuum value indicating a large component of the nucleon mass that is not associated
with the spontaneous breaking of chiral symmetry.
We discuss its connection to the chiral-invariant mass $m_0^{}$ that figures in the
parity-doublet baryon model.
\end{abstract}


\maketitle


\section{Introduction}
\label{sec:intro}

The properties of hadronic matter at high density are poorly understood both theoretically and
experimentally, and pose a challenge in nuclear and particle physics.
They are critically concerned with such issues as the equation of state (EoS) relevant for
compact-star matter and the chiral symmetry breaking/restoration in dense matter.

The sign problem in lattice QCD and the nonperturbative nature of the strong interaction 
in the effective field theory approaches restrict their applicability for dense matter studies.
A series of  works~\cite{PV09} have shown that the skyrmion approach, where the classical 
soliton solutions of the mesonic theory capturing quantum chromodynamics (QCD) in 
the large $N_c$ limit are interpreted as baryons, provides a natural framework for 
exploring dense baryonic matter without being obstructed by the notorious problems 
mentioned above.
Furthermore, it enables one to study the properties of dense matter and in-medium 
hadrons in a unified way~\cite{LPMRV03}. 
However, because of the parameter dependence in the results of simple models that 
include only a few low-lying mesons, the skyrmion approach could not provide 
quantitatively meaningful predictions.

Recently, we have studied the skyrmion approach~\cite{MOYHLPR12,MYOH12,MHLOPR13,MYOHLPR13}
using a chiral Lagrangian, where the $\rho$ and $\omega$ mesons are introduced as
the gauge bosons of the hidden local symmetry (HLS)~\cite{BKUY85, BKY88, HY03a}
with the $O(p^4)$ terms in the chiral order expansion including all the homogeneous
Wess-Zumino (hWZ) terms taken into account. 
All the parameters of the Lagrangian are fixed by their relations to the 
Sakai-Sugimoto's five-dimensional holographic QCD (hQCD) model~\cite{SS04a-SS05}, 
while two parameters of the latter are fixed to yield the empirical values of the 
pion decay constant and the vector meson mass in matter-free space.
We shall refer to this model as HLS($\pi,\rho,\omega$).

Several remarkable results, qualitatively different from, or not observed in, the previous 
works on skyrmions were obtained in Refs.~\cite{MOYHLPR12,MYOH12,MHLOPR13,MYOHLPR13}.
We will return to some of them later. 
Here we mention a few to illustrate the key issues that will concern what we 
treat in this paper.

\begin{itemize}

\item[(i)]
Although the results show some discrepancies from nature, if one considers the fact 
that none of the parameters is adjusted with baryon properties, they could be taken 
as the first ``parameter-free" predictions of the skyrmion approach. 
Furthermore, in spite of the limitations of the model such as the large $N_c$, 
large 't Hooft constant ($\lambda$), and chiral limits taken in the hQCD model, the
semiquantitative agreement of the results with experiments is quite remarkable.

\item[(ii)]
The full $O(p^4)$ terms, in particular the hWZ terms that carry the $\omega$ meson
degree of freedom, are found to be essential for the soliton structure of elementary 
baryons.
This highlights the particularly important role, thus far unexposed, played by the 
$\omega$ meson in the baryon structure. 
We assert that the Lagrangian given up to $O(p^4)$ cannot be approximated by a few terms
as was done in the past.%
\footnote{These terms include the quartic Skyrme term in the pion-only chiral 
Lagrangian as in the original Skyrme model or the ``minimal" HLS model with one hWZ 
term out of the three (explained below).}

\item[(iii)]
In dense matter simulated on the face-centered cubic (FCC) crystal, it is found that
the transition from skyrmions to half-skyrmions accompanied by the vanishing quark
condensate takes place near the normal nuclear matter density (denoted by $n_0^{}$
from here on) in contrast to the case without the vector mesons and the $O(p^4)$ terms.
In the latter case, the transition density, denoted $n_{1/2}^{}$, is much too high compared
with $n_0^{}$. 
A qualitatively different feature found there, which has not been noticed in the past, 
is that the medium-modified pion decay constant $f_\pi^*$ that decreases smoothly
up to $n_{1/2}^{}$ roughly in the same way as in chiral perturbation calculations
stops decreasing at $n_{1/2}^{}$ and then stays more or less constant, although
the space averaged quark condensation $\langle \bar{q}q \rangle$ vanishes in the
half-skyrmion phase. 
The nonvanishing $f_\pi^*$ in the half-skyrmion phase implies that the chiral 
symmetry remains still in the Nambu-Goldstone mode even though 
$\langle \bar{q}q\rangle = 0$.
Thus, the quark condensate is not a good order parameter in the crystal description.

\item[(iv)]
It is found that the medium-modified effective nucleon mass $m_N^\ast$ tracks closely
$f_\pi^\ast$, indicating that the large $N_c$ approximation continues to hold in medium.
Furthermore, in the half-skyrmion phase, the mass remains $\sim 60$ \% of its vacuum value.
The condensate-independent mass is reminiscent of the chiral-invariant baryon mass 
$m_0^{}$ that figures in the parity-doublet baryon model~\cite{DK89}, but its physical origin is not clarified yet.

\end{itemize}

We note that some of the features mentioned above are in disagreement with what was 
obtained in the mean field approximation~\cite{BR91} and in a naive formulation of 
HLS in the meson sector~\cite{HY03a}. 
In particular, they bring tension with the vector manifestation (VM)%
\footnote{In the formulation of the VM in Ref.~\cite{HY03a}, the vanishing quark condensate
($\langle \bar{q}q \rangle=0$) was used. Of course the vanishing of the $\rho$ meson mass
can never occur unless $f_\pi=0$ at the chiral restoration point identified with the VM.}
and the Brown-Rho (BR) scaling.

Given that what we are dealing with here is an effective theory of QCD given 
in terms of the ``macroscopic" degrees of freedom, i.e., hadrons, we need to match 
the effective theory to QCD at a certain scale at which the chiral symmetry gets restored.
By matching the correlators of HLS to those of QCD in the Wilsonian sense, one finds 
that near the chiral restoration scale, here the putative critical density $n_c^{}$, 
the parameters in HLS should satisfy the following relations:
\begin{eqnarray}
f_\pi^2 (q^2=0) \to 0, \quad m_\rho^2 \to m_\pi^2 = 0, \quad
a(q^2=m^2_\rho) \to 1, \nonumber\\
\label{eq:VM}
\end{eqnarray}
which are called ``vector manifestation" of the Wigner realization of chiral symmetry~\cite{HY00}.
The matching between the effective theory and QCD renders the low energy constants (LECs) of
the theory \textit{intrinsically density dependent\/}.%
\footnote{It should be stressed that when we refer to density dependence, we are referring
specifically to this \textit{intrinsic density dependence\/}~\cite{BR03}.
Any truncation in a many-body system in the sense of a Wilsonian renormalization group (RG)
would generate density dependence in the parameters of the model, and it would be in practice
very difficult to identify or isolate the intrinsic density dependence coming from the matching to
QCD in nuclear observables. This has compounded the efforts to see ``partial chiral symmetry
restoration" from nuclear experiments.}
However, if, as mentioned, $f_\pi^\ast$ stays as a nonzero constant in the half-skyrmion 
phase and $m_\rho^{}$ does not scale in medium, it is not clear how to access the VM 
unless some other (hadronic) mechanism intervenes before the QCD degrees of freedom enter.

The objective of this paper is to resolve this problem.
For this, we resort to scale (or conformal) symmetry of QCD.

It is well known that the chiral symmetry breaking and scale symmetry are intricately 
linked to each other and that chiral symmetry breaking could be triggered by spontaneous 
breaking of scale symmetry which in turn is caused by the presence of explicit scale symmetry
breaking~\cite{FN68,LR13b}.
This is the mechanism that we exploit to investigate the possible realization of VM by 
incorporating the QCD trace anomaly in HLS. 
We find that, with an appropriate inclusion of the scalar field associated with the 
explicit breaking of conformal symmetry, both VM and the BR scaling~\cite{BR91} 
can be realized.

There are a variety of reasons to believe that the scalar degree of freedom is needed 
in addition to those that figure in HLS in the structure of both elementary baryon 
and multibaryon systems.
Firstly the skyrmion mass obtained in HLS($\pi,\rho,\omega$)~\cite{MOYHLPR12,MYOH12} 
overshoots the empirical nucleon mass by $\gtrsim 300$~MeV.
One expects that the $O(N_c^0)$ Casimir energy, missing in the $O(N_c)$ soliton mass, 
can account for the attraction needed of that amount.
In the Skyrme model, the Casimir contribution comes from pion loops that account 
for the scalar channel and is of the right magnitude to lower the mass from the 
canonical $\sim 1500$~MeV down to $\sim 1000$~MeV~\cite{MW96}.
Secondly, in the mean field approach to nuclear matter, a (chiral) scalar mass of 
$\sim 600$~MeV is indispensable for binding the matter, counterbalancing the 
repulsion from the $\omega$ meson exchange.
Closer to the problem at hand, the simulation of the nuclear property from the FCC 
crystal shows that the density $n_{\rm min}^{}$ at which the per-skyrmion energy is 
the minimum is larger than the normal nuclear density $n_0^{}$ and the binding 
energy at $n_{\rm min}^{}$ is $\sim 100$~MeV, which is larger than the empirical 
value $\sim 16$~MeV per baryon~\cite{MHLOPR13}.
These discrepancies also signal that there is something missing in the present 
skyrmion crystal description for the EoS of nuclear matter in HLS($\pi,\rho,\omega$). Although there is such a defect in our approach, in fact, what we are suggesting here is that one can make predictions on certain fluctuation properties of hadrons on top of the skyrmion background which is treated semiclassically, which is what the simulation all about.

The advantage of having the vector mesons as (hidden) gauge fields is that a systematic 
chiral perturbation theory can be formulated with vector mesons in addition to the 
pion~\cite{MHLOPR13}.%
\footnote{The vector manifestation cannot be obtained unless the vector meson mass can be
considered as light as the pion mass as in HLS. The phenomenological Lagrangians used in
the literature for treating vector mesons in dense medium can make sense only in mean field
and cannot address dropping vector meson masses.}
Introducing scalar fields in this framework is, however, highly problematic.
What was done in the past and will be done here is to use the trick of the conformal 
compensator field or ``conformon"%
\footnote{This term is borrowed from Ref.~\cite{KL13} in which it is used in cosmology.}
used in cosmology and also in the technidilaton approach to a Higgs-like boson, to write 
a conformally invariant Lagrangian with the conformal symmetry spontaneously broken 
by a Coleman-Weinberg-type potential. 
The conformon field is then identified as a (pseudo-)Nambu-Goldstone boson of 
spontaneously broken scale symmetry, i.e., the dilaton.

In what follows we analyze in what way this dilaton resolves the problem mentioned 
above. This will reveal how it affects the structure of both the elementary nucleon and 
multibaryon systems at large density.

This paper is organized as follows.
In Sec.~\ref{sec:sumhls}, after a sketch of the general strategy of approaching 
elementary baryon and multibaryon systems with one single Lagrangian anchored on 
hidden local symmetry, we introduce the dilaton associated with the scale symmetry 
breaking of QCD as a conformon to HLS. 
This section serves also to define the notations used in the present work.
In Sec.~\ref{sec:single} the single skyrmion properties are studied using the dilaton
compensated HLS model.
The effects of the dilaton on the skyrmion matter properties and medium-modified
hadron properties are explored in Sec.~\ref{sec:matter}.
A possible way to realize VM and restore chiral symmetry with the effects of 
dilaton is also discussed.
We give a succinct summary of the results in Sec.~\ref{sec:summary} and further
discussions in Sec.~\ref{sec:dis}.


\section{Hidden Local Symmetry Lagrangian with Conformal Invariance }

\label{sec:sumhls}

We start with a brief description of the HLS Lagrangian for defining the notations 
used in this paper.
In free space, the full symmetry group associated with the basic ingredients, $\pi$, 
$\rho$, and $\omega$, is $G_{\rm full} = [ \mbox{SU(2)}_L \times \mbox{SU(2)}_R ]_{\rm chiral} 
\times [\mbox{U(2)}]_{\rm HLS}$ in which the lowest-lying $\rho$ and $\omega$ mesons 
are incorporated as the gauge bosons of $[\mbox{SU(2)}]_{\rm HLS}$ and 
$[\mbox{U(1)}]_{\rm HLS}$ components, respectively, of the ``hidden local symmetry'' 
$[\mbox{U(2)}]_{\rm HLS}$.
The HLS Lagrangian is constructed by two 1-forms, $\hat{\alpha}_{\parallel \mu}^{}$ and
$\hat{\alpha}_{\perp \mu}^{}$, defined by
\begin{eqnarray}
\hat{\alpha}_{\parallel \mu}^{} & = &
\frac{1}{2i} (D_\mu \xi_R^{} \cdot \xi_R^\dagger + D_\mu \xi_L^{}\cdot \xi_L^\dagger),
\label{eq:defalphapara}\\
\hat{\alpha}_{\perp \mu}^{} & = &
\frac{1}{2i} (D_\mu \xi_R^{}\cdot \xi_R^\dagger - D_\mu \xi_L^{}\cdot \xi_L^\dagger),
\label{eq:defalphaperp}
\end{eqnarray}
with the chiral fields $\xi_L^{}$ and $\xi_R^{}$, which are expressed in the unitary 
gauge as
\begin{eqnarray}
\xi_L^\dagger & = & \xi_R^{} = e^{i \pi/2f_\pi} \equiv \xi
\quad \mbox{with} \quad \pi = \bm{\pi} \cdot \bm{\tau},
\end{eqnarray}
where $\bm{\tau}$'s are the Pauli matrices.
The covariant derivative associated with the hidden local symmetry is defined as
\begin{eqnarray}
D_\mu \xi_{R,L}^{} &=& (\partial_\mu - i V_\mu) \, \xi_{R,L}^{}, 
\label{eq:covDhls}
\end{eqnarray}
where $V_\mu$ represents the gauge boson of the HLS~\cite{BKUY85,BKY88,HY03a} as%
\footnote{In this paper, we distinguish the gauge coupling constants for the $\omega$ 
and the $\rho$ mesons which will be convenient for discussing the medium modified 
hadron properties. In free space, we take the hidden gauge symmetry as 
$\mbox{U(2)}_{\rm HLS}$, thus $g_\omega = g_\rho \equiv g$. 
If $\mbox{U(2)}_{\rm HLS}$ is broken in dense medium, they could have different values.}
\begin{eqnarray}
V_\mu = \frac{1}{2} \left( g_\omega^{} \omega_\mu^{} + g_\rho^{} \rho_\mu^{} \right)
\end{eqnarray}
and
\begin{eqnarray}
\rho_\mu^{} = \bm{\rho}_\mu^{} \cdot \bm{\tau} =
\left( \begin{array}{cc}
  \rho_\mu^0  & \sqrt{2} \rho_\mu^+ \\
  \sqrt{2} \rho_\mu^- &  -\rho_\mu^0
\end{array} \right) .
\end{eqnarray}

Up to $O(p^4)$, including the hWZ terms, the most general HLS Lagrangian can be 
expressed as
\begin{eqnarray}
\mathcal{L}_{\rm HLS} & = &
\mathcal{L}_{(2)}^{\rm HLS} + \mathcal{L}_{(4)}^{\rm HLS} + 
\mathcal{L}_{\rm anom}^{\rm HLS} ,
\label{eq:Lag_HLS}
\end{eqnarray}
with
\begin{eqnarray}
\mathcal{L}_{(2)}^{\rm HLS} &=&
f_\pi^2 \,\mbox{Tr}\, 
\left( \hat{\alpha}_{\perp\mu}^{} \hat{\alpha}_{\perp}^{\mu} \right)
+ a f_\pi^2 \,\mbox{Tr}\, \left(\hat{\alpha}_{\parallel\mu}^{} 
\hat{\alpha}_{\parallel}^{\mu} \right)
+ \mathcal{L}_{\rm kin},
\end{eqnarray}
where $f_\pi$ is the pion decay constant, $a$ is the HLS parameter,
and $\mathcal{L}_{\rm kin}$ contains the kinetic terms of the vector mesons:
\begin{eqnarray}
\mathcal{L}_{\rm kin} &=&
- \frac{1}{2g_\rho^2} \mbox{Tr}\, \left( V_{\mu\nu}^{(\rho)} V^{(\rho), \mu\nu} \right) 
\nonumber\\
& & \mbox{} - \frac{1}{2g_\omega^2} \mbox{Tr}\, \left( V_{\mu\nu}^{(\omega)}
V^{(\omega), \mu\nu} \right),
\label{eq:Lkin}
\end{eqnarray}
with the field-strength tensors of vector mesons
\begin{eqnarray}
V_{\mu\nu}^{(\rho)} &=& 
\partial_\mu \left( \textstyle\frac{1}{2} g_\rho^{} \rho_\nu^{} \right)
- \partial_\nu \left( \textstyle\frac{1}{2} g_\rho^{} \rho_\mu^{} \right) 
- i \left[ \textstyle \frac{1}{2} g_\rho^{} \rho_\mu^{} , 
\textstyle\frac{1}{2} g_\rho^{} \rho_\nu^{} \right] , 
\nonumber\\
V_{\mu\nu}^{(\omega)} &=& 
\partial_\mu \left( \textstyle\frac{1}{2} g_\omega^{} \omega_\nu^{}\right) -
\partial_\nu \left( \textstyle\frac{1}{2} g_\omega^{} \omega_\mu^{} \right).
\end{eqnarray}
For later discussions, we have separated the terms for the $\rho$ and $\omega$ mesons
to allow different values for $g_\rho^{}$ and $g_\omega^{}$ which are the same
in the case of $\mbox{[U(2)]}_{\rm HLS}$.

The $O(p^4)$ Lagrangian is given by~\cite{HY03a,Tana93}
\begin{widetext}
\begin{eqnarray}
\mathcal{L}_{(4)} &=&
y_1^{} \mbox{Tr} \Bigl[ \hat{\alpha}_{\perp\mu}^{} \hat{\alpha}_\perp^\mu
\hat{\alpha}_{\perp\nu}^{} \hat{\alpha}_\perp^\nu \Bigr]
+ y_2^{} \mbox{Tr} \Bigl[ \hat{\alpha}_{\perp\mu}^{} \hat{\alpha}_{\perp\nu}^{}
\hat{\alpha}^\mu_\perp \hat{\alpha}^\nu_\perp \Bigr]
+ y_3^{} \mbox{Tr}
\left[ \hat{\alpha}_{\parallel\mu}^{} \hat{\alpha}_\parallel^\mu
\hat{\alpha}_{\parallel\nu}^{} \hat{\alpha}_\parallel^\nu \right]
+ y_4^{} \mbox{Tr}
\left[ \hat{\alpha}_{\parallel\mu}^{} \hat{\alpha}_{\parallel\nu}^{}
\hat{\alpha}^\mu_\parallel \hat{\alpha}^\nu_\parallel \right]
\nonumber \\ && \mbox{}
+ y_5^{} \mbox{Tr}
\left[ \hat{\alpha}_{\perp\mu}^{} \hat{\alpha}_\perp^\mu
\hat{\alpha}_{\parallel\nu}^{} \hat{\alpha}_\parallel^\nu \right]
+ y_6^{} \mbox{Tr}
\left[ \hat{\alpha}_{\perp\mu}^{} \hat{\alpha}_{\perp\nu}^{}
\hat{\alpha}^\mu_\parallel \hat{\alpha}^\nu_\parallel \right]
+ y_7^{} \mbox{Tr}
\left[ \hat{\alpha}_{\perp\mu}^{} \hat{\alpha}_{\perp\nu}^{}
\hat{\alpha}^\nu_\parallel \hat{\alpha}^\mu_\parallel \right]
\nonumber \\ && \mbox{}
+ y_8^{} \left\{
\mbox{Tr} \left[ \hat{\alpha}_{\perp\mu}^{} \hat{\alpha}_\parallel^\mu
\hat{\alpha}_{\perp\nu}^{} \hat{\alpha}_\parallel^\nu \right]
+ \mbox{Tr} \left[ \hat{\alpha}_{\perp\mu}^{} \hat{\alpha}_{\parallel\nu}^{}
\hat{\alpha}_\perp^\nu \hat{\alpha}_\parallel^\mu \right] \right\}
+ y_9^{} \mbox{Tr}
\left[ \hat{\alpha}_{\perp\mu}^{} \hat{\alpha}_{\parallel\nu}^{}
\hat{\alpha}^\mu_\perp \hat{\alpha}^\nu_\parallel \right]
\nonumber\\
 & & \mbox{} +
i z_4^{} \mbox{Tr}
\Bigl[ V_{\mu\nu}^{(\rho)} \hat{\alpha}_\perp^\mu \hat{\alpha}_\perp^\nu \Bigr]
+ i z_5^{} \mbox{Tr}
\left[ V_{\mu\nu}^{(\rho)} \hat{\alpha}_\parallel^\mu \hat{\alpha}_\parallel^\nu \right].
\end{eqnarray}
\end{widetext}
Note that $V_{\mu\nu}^{(\omega)}$ does not appear in the $z_4^{}$ and $z_5^{}$ terms.

Finally, the anomalous parity hWZ terms $\mathcal{L}_{\rm anom}$ are written as
\begin{eqnarray}
\Gamma_{\rm hWZ} = \int d^4x \mathcal{L}_{\rm anom} = \frac{N_c}{16\pi^2}\int_{M^4}
\sum_{i=1}^3 c_i \mathcal{L}_i ,
\end{eqnarray}
where $M^4$ stands for the four-dimensional Minkowski space and
\begin{subequations}
\begin{eqnarray}
\mathcal{L}_1 & = & i \, \mbox{Tr}\,
\bigl[ \hat{\alpha}_{L}^3 \hat{\alpha}_{R}^{}
 - \hat{\alpha}_{R}^3 \hat{\alpha}_{L}^{} \bigr], \\
\mathcal{L}_2 & = & i \, \mbox{Tr}\,
\bigl[ \hat{\alpha}_{L}^{} \hat{\alpha}_{R}^{}
\hat{\alpha}_{L}^{} \hat{\alpha}_{R}^{} \bigr]  ,  \\
\mathcal{L}_3 & = & \mbox{Tr}\,
\bigl[ F_{V} \left( \hat{\alpha}_{L}^{} \hat{\alpha}_{R}^{}
 - \hat{\alpha}_{R}^{} \hat{\alpha}_{L}^{} \right) \bigr] ,
\end{eqnarray}
\end{subequations}
in the 1-form and 2-form notations with
\begin{eqnarray}
\hat{\alpha}_{L}^{} & = & \hat{\alpha}_\parallel^{} - \hat{\alpha}_\perp^{},
\nonumber \\
\hat{\alpha}_{R}^{} & = & \hat{\alpha}_\parallel^{} + \hat{\alpha}_\perp^{},
\nonumber \\
F_V & = & dV - i V^2.
\end{eqnarray}

In the Lagrangian (\ref{eq:Lag_HLS}) there appear many undetermined constants 
which include $f_\pi$, $a$, $g_\rho^{}$, $g_\omega^{}$, $y_i^{} (i=1,\cdots,9)$, 
$z_i^{} (i=4,5)$, and $c_i^{} (i=1,2,3)$.
To fix them phenomenologically, we need a large number of experimental data, 
which are not available at present and will not be available in the near future.
The recent development of holographic QCD, however, improves the situation 
dramatically.
As discussed in Refs.~\cite{MOYHLPR12,MYOH12}, those coefficients can be fixed 
completely by means of a set of  ``master formulas"  that  match the 
four-dimensional effective theory (here HLS) to the five-dimensional hQCD model.
In the large $N_c$ and large $\lambda$ limit, the hQCD has two parameters which 
can be related to the empirical values of the pion decay constant and the vector 
meson mass.
Then with these quantities fixed in the meson sector, all the coefficients of the 
HLS Lagrangian we are dealing with are determined by the master formula.
Here we employ the Sakai-Sugimoto hQCD model~\cite{SS04a-SS05} which is
supposed to be dual to our HLS model, with the empirical values
\begin{eqnarray}
f_\pi = 92.4 \mbox{ MeV}, \qquad m_\omega^{} = m_\rho^{} = 775.5 \mbox{ MeV},
\label{eq:PMS}
\end{eqnarray}
where the $\mbox{[U(2)]}_{\rm HLS}$ in free space has been taken.

The essential point in deriving HLS from hQCD models that have 5D Dirac-Born-Infeld
part and the Chern-Simons part,
\begin{eqnarray}
S_{\rm 5} & = & S_{\rm 5}^{\rm DBI} + S_{\rm 5}^{\rm CS},
\label{eq:actionhQCD}
\end{eqnarray}
where
\begin{eqnarray}
S_{\rm 5}^{\rm DBI} & = & N_c G_{\rm YM} \int d^4 x dz
\bigg\{ - \frac{1}{2} K_1(z) \mbox{Tr}
  \left[ \mathcal{F}_{\mu\nu} \mathcal{F}^{\mu\nu} \right]
\nonumber\\
&& \qquad \qquad  \mbox{}
+ K_2(z) M_{KK}^2 \mbox{Tr} \left[ \mathcal{F}_{\mu z} \mathcal{F}^{\mu z}
\right]
\biggr\},
\label{eq:DBI} \\
S_{\rm 5}^{\rm CS} & = & \frac{N_c}{24\pi^2} \int_{M^4\times R} w_5^{} (A),
\label{eq:SSCS}
\end{eqnarray}
is to make the mode expansion of the 5D gauge field $A_M(x,z)$ [$M = (\mu, z)$ 
with $\mu = 0, 1, 2, 3$] and integrate out all the modes except the pseudoscalar 
and the lowest-lying vector mesons.
This reduces $A_M(x,z)$ to $A_M^{\rm integ}(x,z)$ which in the $A_z(x,z)= 0$ gauge 
amounts to the substitution~\cite{MOYHLPR12,MYOH12}
\begin{eqnarray}
A_\mu(x,z) &\to& A_\mu^{\rm integ}(x,z) 
\nonumber\\
&=& \hat{\alpha}_{\perp \mu}^{} \psi_0 
+ \left( \hat{\alpha}_{\parallel \mu}^{} + V_\mu\right) +
\hat{\alpha}_{\parallel \mu}^{} \psi_1(z), 
\label{eq:Ainteu2}
\end{eqnarray}
where $\{\psi_n(z)\}$ are the eigenfunctions satisfying the following eigenvalue 
equation obtained from the action:
\begin{eqnarray}
- K_1^{-1}(z)\partial_z \left[ K_2(z) \partial_z \psi_n^{} (z) \right]
= \lambda_n^{} \psi_n^{}(z), \label{eq:eomhqcd}
\end{eqnarray}
with $\lambda_n$ being the $n$th eigenvalue ($\lambda_0 = 0$).
Here, $K_1(z)$ and $K_2(z)$ are the warping factors in the fifth direction of the 
five-dimensional space-time, which are explicitly $K_1(z) = K^{-1/3}(z)$ and 
$K_2(z) = K(z)$ with $K(z) = 1 + z^2$ in the Sakai-Sugimoto model~\cite{SS04a-SS05}.

In this paper, in order to distinguish the $\rho$ and $\omega$ mesons associated 
to the $\mbox{SU(2)}$ and $\mbox{U(1)}$ components, respectively, of the 5D gauge 
field $A_M(x,z)$, we rewrite Eq.~(\ref{eq:Ainteu2}) as
\begin{eqnarray}
A_\mu(x,z) &\to& A_\mu^{\rm integ}(x,z) 
\nonumber\\
&=& \hat{\alpha}_{\perp \mu}^{} \psi_0 
+ \left(\hat{\alpha}_{\parallel \mu}^{} + V_\mu\right) 
\nonumber\\
&& \mbox{} + \hat{\alpha}_{\parallel \mu}^{\rm SU(2)} \psi_1(z)
+ \hat{\tilde{\alpha}}_{\parallel \mu}^{\rm U(1)} \tilde{\psi}_1(z), 
\label{eq:Aintesu2u1}
\end{eqnarray}
where again we have separated out the SU(2) part and the U(1) part in the last 
two terms.
The expression for $\hat{\tilde{\alpha}}_{\parallel \mu}^{\rm U(1)}$ can be 
obtained from Eq.~\eqref{eq:defalphapara} by removing all the isotriplets.
With these conventions one can easily see that $\psi_1(z)$ and $\tilde{\psi}_1(z)$ 
are the wave functions of the $\rho$ and $\omega$ mesons, respectively.

Substituting Eq.~\eqref{eq:Aintesu2u1} into the five-dimensional hQCD models leads 
to the HLS Lagrangian.
For the $O(p^2)$ terms we have the following relations for the low energy constants:
\begin{eqnarray}
f_{\pi}^2 & = & N_c G_{\rm YM}^{} M_{KK}^2  \int dz K_2(z) \left[ \dot{\psi}_0^{}(z)
\right]^2, \nonumber\\
a f_{\pi}^2 & = & N_c G_{\rm YM}^{} M_{KK}^2 \lambda_1^{} \langle \psi^2_1 \rangle,
\nonumber\\
\frac{1}{g_\rho^2} & = & N_c G_{\rm YM}^{} \langle \psi_1^2 \rangle ,
\nonumber\\
\frac{1}{g_\omega^2} & = & N_c G_{\rm YM}^{} \langle \tilde{\psi}_1^2 \rangle .
\end{eqnarray}
The parameters $a$, $f_\pi$, and the gauge coupling constants $g_\rho^{}$ and 
$g_\omega^{}$ satisfy the following relations:
\begin{eqnarray}
a g_\rho^2 f_\pi^2 = m_\rho^2, \qquad a g_\omega^2 f_\pi^2 = m_\omega^2.
\end{eqnarray}
Therefore, $\psi_1(z)$ and $\tilde{\psi}_1(z)$ satisfy
\begin{eqnarray}
\tilde{\psi}_1(z) = \frac{m_\rho^{}}{m_\omega^{}}\psi_1(z) .
\label{eq:relationpsi}
\end{eqnarray}
By using Eq.~\eqref{eq:relationpsi}, we get the master formula of the low energy 
constants of the $O(p^4)$ terms as~\cite{MasterFormula}%
\footnote{Here, we express the master formula in terms of $f_\pi$ and $m_\rho^{}$.
The factor $m_\rho^{}/m_\omega^{}$ in $c_i^{}$ arises from the normalization of 
$\psi_1$.
In free space we have $m_\rho^{}/m_\omega^{} = 1$.}
\begin{eqnarray}
y_1^{} & = & -y_2^{} = - \frac{f_\pi^2}{m_\rho^2}N_{\rm hQCD} \left\langle
\left(1 + \psi_1 - \psi_0^2 \right)^2
\right\rangle ,
\nonumber\\
y_3^{} & = & -y_4^{} = - \frac{f_\pi^2}{m_\rho^2} N_{\rm hQCD} \left\langle
\psi^2_1 \left(1 + \psi_1^{} \right)^2
\right\rangle ,
\nonumber\\
y_5^{} & = & 2 y_8^{} = -y_9^{} = -2\frac{f_\pi^2}{m_\rho^2}N_{\rm hQCD}
\left\langle \psi_1^2 \psi_0^2 \right\rangle ,
\nonumber\\
y_6^{} & = & - \left( y_5^{} + y_7^{} \right) ,
\nonumber\\
y_7^{} & = & \frac{2f_\pi^2}{m_\rho^2}N_{\rm hQCD}  \left\langle \psi_1^{}
\left ( 1 + \psi_1^{} \right)
\left(1 + \psi_1^{} - \psi_0^2 \right) \right\rangle ,
\nonumber\\
z_4^{} & = & \frac{2f_\pi^2}{m_\rho^2} N_{\rm hQCD} \left\langle \psi_1^{}
\left( 1+\psi_1^{} - \psi_0^2 \right)
\right\rangle ,
\nonumber\\
z_5^{} & = & - \frac{2f_\pi^2}{m_\rho^2} N_{\rm hQCD} \left\langle \psi_1^2
\left( 1 + \psi_1^{} \right) \right\rangle ,
\nonumber\\
c_1^{} & = &  \frac{m_\rho}{m_\omega} \left\langle\hskip -0.5em \left\langle
\dot{\psi}_0^{} \psi_1^{} \left( \frac{1}{2} \psi_0^2 + \frac{1}{6} \psi_1^2
- \frac{1}{2} \right) \right\rangle\hskip -0.5em \right\rangle ,
\nonumber\\
c_2^{} & = & \frac{m_\rho}{m_\omega} \left\langle\hskip -0.5em \left\langle
\dot{\psi}_0^{} \psi_1^{} \left( -\frac{1}{2} \psi_0^2 + \frac{1}{6} \psi_1^2
+ \frac{1}{2} \psi_1^{} + \frac{1}{2} \right) \right\rangle\hskip -0.5em \right\rangle,
\nonumber\\
c_3^{} & = & \frac{m_\rho}{m_\omega} \left\langle\hskip -0.5em \left\langle
\frac{1}{2}\dot{\psi}_0^{} \psi_1^{2} \right\rangle\hskip -0.5em \right\rangle ,
\label{eq:lecshls}
\end{eqnarray}
where $N_{\rm hQCD}  = \lambda_1/\int dz K_2(z)[\dot{\psi}_0(z)]^2$, and the 
wave function $\tilde{\psi}_1(z)$ associated with the $\omega$ field in $c_i^{}$ 
has been expressed in terms of $\psi_1(z)$ associated with the $\rho$ field 
through the relation (\ref{eq:relationpsi}).
For deriving the expressions of $y_i^{}$ and $z_i^{}$, we have considered that the 
U(1) degree of freedom should disappear in these terms because of the antisymmetric 
field tensor appearing in the Dirac-Born-Infeld part. 
The integrals appearing in the above relations are defined by
\begin{eqnarray}
\langle A \rangle &\equiv& \int_{-\infty}^{\infty}  dz K_1(z) A(z),
\nonumber\\
\langle\hskip -0.2em \langle A \rangle\hskip - 0.2em \rangle 
&\equiv & \int_{-\infty}^\infty dz A(z).
\end{eqnarray}

\subsection{Predictions of HLS}

To highlight the principal effect of the dilaton in dense skyrmion matter, we briefly 
review the predictions of dilatonless HLS obtained in the previous works.
In Ref.~\cite{MYOH12} it was shown that this model yields the soliton mass of 
$1184$~MeV, which is quite good as a ``parameter-free" result.
Although it is larger by about $300$~MeV than the observed nucleon mass, it is not 
difficult to understand where this difference may come from.
As mentioned in Sec.~\ref{sec:intro}, in the standard Skyrme model (with pions only), 
an excess of $\sim 500$~MeV can be reduced by the Casimir energy~\cite{MW96} that 
comes at the next order in $N_c$, i.e., $O(N_c^0)$. 
We will see below that the dilaton contributes to remove \textit{a part} of that 
excess, although not enough.

The HLS Lagrangian also provides a noticeable improvement in the dense baryonic 
matter study~\cite{MHLOPR13} compared to what exists in the literature.
Here, the skyrmion--half-skyrmion transition takes place near the normal nuclear 
matter density, rendering the process phenomenologically relevant.
A distinctively novel result is that in the half-skyrmion phase, the intrinsic 
density-dependent (or effective in-medium) pion decay constant $f_\pi^{\ast}$ is 
nonvanishing and stays independent of density.
The in-medium nucleon mass $m_N^{\ast}$ scales similarly to the pion decay constant, which is
indicative of the large $N_c$ dominance.
This (nearly) constant nucleon mass in the half-skyrmion phase resembles the 
nonvanishing chiral-invariant mass in the parity-doublet chiral model for 
baryons~\cite{DK89}.
We will return to this matter in the discussion section in connection with the 
origin of the nucleon mass.

What seems not evident is the movement toward the vector manifestation of the HLS 
given in Eq.~(\ref{eq:VM}). 
In order to arrive at the fixed point that would correspond to the chiral transition,
which is a quantum phase transition, the correlators of HLS should be matched to 
those of QCD. 
For this, it is clear that one has to understand the quantum structure of the 
half-skyrmion phase. 
As suggested in Ref.~\cite{LR13c}, it could involve a topology-triggered change 
from a Fermi-liquid state to a non-Fermi-liquid state. 
This issue needs to be clarified.

\subsection{Conformally compensating HLS}

It is also plausible that higher-order corrections and/or heavier vector mesons 
such as the $a_1^{}$ could play an important role in approaching the chiral 
restoration point.
Our thesis in this paper, as stated in Sec.~\ref{sec:intro}, is that what is crucially 
needed in the HLS structure is the scalar degree of freedom.

As stated in Sec.~\ref{sec:intro}, we introduce the scalar needed as a dilaton that
figures in spontaneous breaking of scale symmetry (SBSS) which is locked to 
spontaneous breaking of chiral symmetry (SBCS)~\cite{LR13b}.
The idea is that the trace anomaly of QCD provides the explicit breaking of 
scale symmetry that is needed to trigger the SBSS.
It is well known that, without the explicit breaking, the spontaneous breaking 
cannot occur~\cite{FN68}.
We associate the part of the gluon condensate that remains ``unmelted" above the 
critical temperature or density, i.e., the ``hard glue" in the language of 
Ref.~\cite{BR03}, with the explicit breaking of scale symmetry.
We follow the standard procedure of incorporating the nonlinearly realized scale 
invariance of adding a field $\chi$ as the ``conformal compensator" (or conformon 
for short).
The procedure is to make the HLS Lagrangian conformally invariant and then add a 
potential $\mathcal{V}$ that breaks conformal invariance spontaneously.
The spontaneous breaking makes the conformon a (pseudo-)Nambu-Goldstone boson, 
i.e., the dilaton.

If one assumes that the vector fields have scale dimension $1$,%
\footnote{This can be supported by the observation that
$$
\rho_\mu^{} \sim \frac{i}{g_\rho^{}} \left( \partial_\mu^{} \xi_R^{} \xi^\dagger_R
+ \partial_\mu^{} \xi_L^{} \xi^\dagger_L \right),
$$
in the limit that $m_\rho^{} \to \infty$.}
then this conformon trick modifies only the $O(p^2)$ term in Eq.~\eqref{eq:Lag_HLS},
since the $O(p^4)$ terms are scale invariant as they are.
Putting in the dilaton part of the Lagrangian, we have
\begin{eqnarray}
\mathcal{L}_{\rm dHLS\mbox{-}I} & = & \mathcal{L}_{\rm (2)}^{\rm dHLS\mbox{-}I}
+ \mathcal{L}_{\rm (4)}^{\rm HLS} + \mathcal{L}_{\rm anom}^{\rm HLS}
+ \mathcal{L}_{\rm dilaton},
\label{eq:lagrdhlsI}
\end{eqnarray}
where
\begin{eqnarray}
\mathcal{L}_{\rm (2)}^{\rm dHLS\mbox{-}I}  &=&
f_\pi^2 \left( \frac{\chi}{f_\chi} \right)^2
\mbox{Tr} \left[ \hat{\alpha}_{\perp\mu}^{} \hat{\alpha}_{\perp}^{\mu} \right]
+ a f_\pi^2 \left( \frac{\chi}{f_\chi} \right)^2
\mbox{Tr} \left[ \hat{\alpha}_{\parallel\mu}^{} \hat{\alpha}_{\parallel}^{\mu} \right]
\nonumber\\
&& \mbox{}
+ \mathcal{L}_{\rm kin}, 
\label{eq:dhls1p2}\\
\mathcal{L}_{\rm dilaton}  &=&  \frac{1}{2} \partial_\mu \chi \partial^\mu \chi
+ \mathcal{V}.
\label{eq:lagrtrace}
\end{eqnarray}
Here $\mathcal{L}_{\rm kin}$ is the kinetic term of vector mesons as given in 
Eq.~(\ref{eq:Lkin}) and $f_\chi$ ($\neq 0)$ is the vacuum expectation value of 
the field $\chi$.
The potential $\mathcal{V}$ in this system is not known except that it should 
reproduce the ``soft glue" part in the trace anomaly~\cite{BR03}.
If one assumes that the conformal symmetry breaking term is small, then the potential 
takes the familiar Coleman-Weinberg form (see, e.g., Ref.~\cite{GGS07})
\begin{eqnarray}
\mathcal{V}= - \frac{m_\chi^2 f_\chi^2}{4} \left[ \left( \frac{\chi}{f_\chi} \right)^4
\left\{ \ln \left( \frac{\chi}{f_\chi} \right) - \frac{1}{4} \right\}
+ \frac{1}{4} \right].
\label{logpot}
\end{eqnarray}
We shall refer to the Lagrangian (\ref{eq:lagrdhlsI}) with the potential given by
Eq.~(\ref{logpot}) as  dHLS-I($\pi,\rho,\omega$).

As we shall see below, since the conformon couples in the same way to both the 
$\rho$ and the $\omega$ mesons as well as to the derivative of the $U$ field, as 
the dilaton condensate decreases with density, the energy density of the system 
diverges with  increasing density, as found in the minimal model in Ref.~\cite{PRV03}.%
\footnote{The so-called minimal model corresponds to the truncated Lagrangian of 
HLS with the LECs $y_i^{} = z_i^{} = c_3^{} = 0 $ and $c_1^{} = - c_2^{} = 2/3$ in 
Eq.~(\ref{eq:Lag_HLS}) below.
This is gotten by dropping all $O(p^4)$ terms in the Lagrangian except one term
$\propto \omega_\mu B^\mu$ (where $B_\mu$ is the baryon number current) in the hWZ
that results if one substitutes the equation of motion for the $\rho$ with the 
$\rho$ mass set to infinity in the hWZ part of the Lagrangian but not elsewhere.
Obviously this limit precludes \textit{ab initio} a dropping $\rho$ mass and hence 
the VM, that we consider unacceptable. }
This leads to a contradictory situation that as the density increases, the pion decay 
constant and the vector meson mass are forced to increase, instead of decrease.

There are two ways out of this conundrum. 
Both resort to the possible breaking of $\mbox{[U(2)]}_{\rm HLS}$ symmetry for 
the $\rho$ and $\omega$.

One is to implement the observation made in Ref.~\cite{PLRS13} that the symmetry 
is broken in the gauge coupling $g_{\omega} \neq g_{\rho}$.
This resolves the above-mentioned problem in a similar way to what is discussed 
in Ref.~\cite{PLRS13}. 
This matter will be addressed in more detail in the discussion section as it raises 
further issues to be explored.

The other is to break $\mbox{[U(2)]}_{\rm HLS}$ symmetry in the mass term so that 
in medium, while the $\rho$ mass has the VM property, the $\omega$ mass does not, 
which would prevent the increasing repulsion. 
This can be done by applying the conformal compensator to the $\rho$ sector but 
not to the $\omega$ sector. 
We have not yet figured out how to justify this structure in a rigorous way.%
\footnote{See Ref.~\cite{LR09} for the relevance of the Freund-Nambu theorem to 
dense matter problems. There the problem of the $\omega$ mass was not addressed.}
What we have done in this paper is simply not to apply the conformon to the 
$\omega$ mass.
To do this we factor out the $\omega$ mass term in the second term of 
$\mathcal{L}_{(2)}^{\rm HLS}$ and couple $\chi^2$ only to the rest.
We shall refer to this Lagrangian as dHLS-II($\pi,\rho,\omega$).
Then it reads
\begin{eqnarray}
\mathcal{L}_{\rm dHLS\mbox{-}II} &=& \mathcal{L}_{\rm (2)}^{\rm dHLS\mbox{-}II}
+ \mathcal{L}_{\rm (4)}^{\rm HLS}
+ \mathcal{L}_{\rm anom}^{\rm HLS} + \mathcal{L}_{\rm dilaton},
\label{eq:lagrdhlsII}
\end{eqnarray}
where
\begin{eqnarray}
\mathcal{L}_{\rm (2)}^{\rm dHLS\mbox{-}II} &=&
f_\pi^2 \left( \frac{\chi}{f_\chi} \right)^2
\mbox{Tr} \left[ \hat{\alpha}_{\perp\mu}^{} \hat{\alpha}_{\perp}^{\mu} \right]
\nonumber\\
&& \mbox{} +
a f_\pi^2 \left( \frac{\chi}{f_\chi} \right)^2
\mbox{Tr} \left[\hat{\alpha}_{\parallel\mu}^{}
\hat{\alpha}_{\parallel}^{\mu} \right]_{\rm SU(2)}
\nonumber\\
&& \mbox{} +
\frac{1}{2} af_\pi^2 g_\omega^2 \omega_\mu^{} \omega^\mu
+ \mathcal{L}_{\rm kin} . 
\label{eq:LagrdHLSII}
\end{eqnarray}
The rest are the same as in dHLS-I($\pi,\rho,\omega$).
In the second term, the subscript $\mbox{SU(2)}$ denotes that only the isovector 
part is considered and the $\omega$ mass term is factored out.
Note that in free space $\chi / f_\chi = 1$, so that the $\mbox{U(2)}_{\rm HLS}$ 
symmetry is restored.
We are considering the case that the medium breaks $\mbox{[U(2)]}_{\rm HLS}$ 
symmetry in such a way that, while the $\rho$ mass scales in density, the $\omega$ 
mass does not.
This is analogous to the weak scaling of the $\omega$-nucleon coupling in 
Ref.~\cite{PLRS13}, where the symmetry breaking is attributed to the vector coupling.
We shall see that this simple modification resolves the long-standing problem started 
in Ref.~\cite{PRV03} and enables chiral symmetry to be restored and the vector 
symmetry to be manifest at some critical density $n_c^{}$.

The incorporation of the dilaton in dHLS-II($\pi,\rho,\omega$) brings in two 
undetermined constants $f_\chi$ and $m_\chi$.
Since there are no experimental values, we shall just take them as free parameters.
We will present the results obtained with~\cite{PRV03}
\begin{eqnarray}
f_\chi & = & 240 \mbox{ MeV}, \nonumber\\
m_\chi^{} & = &  720 \mbox{ MeV}.
\label{eq:chi_param}
\end{eqnarray}
In the limit of $m_\chi^{} \rightarrow \infty$, all of the numerical results trivially 
converge to those of HLS($\pi,\rho,\omega$) reported in 
Refs.~\cite{MOYHLPR12,MYOH12,MHLOPR13,MYOHLPR13}.


\section{Single skyrmion properties in \lowercase{d}HLS-I and \lowercase{d}HLS-II}
\label{sec:single}

We first study the effect of the dilaton on the single skyrmion properties.
The soliton solution can be found in the spherical form as
\begin{eqnarray}
\xi(\bm{r}) &=& \exp \left[ i {\bm \tau}\cdot\hat{r} F(r)/2 \right], 
\nonumber\\
{\rho}^\mu (\bm{r}) &=& \frac{G(r)}{g_\rho^{} r} 
\left( \hat{\bm{r}} \times \bm{\tau} \right)^i \delta^{\mu i},  
\nonumber \\
\omega^\mu (\bm{r}) &=& W(r) \delta^{\mu 0}, 
\nonumber\\
\chi({\bm r}) &=& f_\chi C(r).
\label{eq:ansatzc}
\end{eqnarray}

The standard collective rotation quantization~\cite{ANW83} is made by the 
transformation
\begin{eqnarray}
\xi(\bm{r}) &\to& \xi(\bm{r},t) = A(t)\, \xi(\bm{r}) A^\dagger(t), 
\nonumber\\
V_{\mu}(\bm{r}) &\to& V_{\mu}(\bm{r},t) 
= A(t)\, V_{\mu}(\bm{r}) A^\dagger(t),
\label{eq:mesoncollective}
\end{eqnarray}
where $A(t)$ is a time-dependent SU(2) matrix, which defines $\bm{\Omega}$ by
\begin{eqnarray}
i \bm{\tau} \cdot \bm{\Omega} & \equiv & A^\dagger(t) \partial_0 A(t),
\label{eq:angularvelocity}
\end{eqnarray}
which leads to the most general forms for the vector-meson excitations as
\begin{eqnarray}
\rho^0 (\bm{r},t) &=& A(t) \frac{2}{g_\rho^{}}
\left[ \bm{\tau} \cdot \bm{\Omega} \, \xi_1^{}(r)
+ \hat{\bm{\tau}} \cdot \hat{\bm{r}} \, \bm{\Omega} \cdot \hat{\bm{r}} \,
\xi_2^{}(r) \right] A^\dagger(t) , 
\nonumber\\
\omega^i (\bm{r},t) &=& \frac{\varphi(r)}{r} 
\left( \bm{\Omega} \times \hat{\bm{r}} \right)^i .
\label{eq:VM_excited}
\end{eqnarray}
Since the dilaton field $\chi$ is a spin-$0$ isoscalar field, it is not affected 
by the collective rotation.
The boundary conditions of the wave functions are given in Ref.~\cite{MYOH12} 
and those for $C(r)$ read
\begin{equation}
C'(0) = 0, \qquad C(\infty) = 1.
\end{equation}

It is then straightforward to calculate the soliton mass and the moment of inertia
from which the equations of motion for the wave functions introduced in
Eqs.~(\ref{eq:ansatzc}) and (\ref{eq:VM_excited}) can be read.
We refer the details to Ref.~\cite{MYOH12}, which can be easily used to obtain
the equations of motion in the case with the dilaton field.


\begin{table*}[t]
\caption{\label{table:skyrmiondhls1}  
Numerical results of the skyrmion properties.
$M_{\rm sol}$ and $\Delta_M (\equiv M_\Delta - M_N)$ are in units of MeV, while
$\sqrt{\langle r^2 \rangle_B}$ and $\sqrt{\langle r^2 \rangle_E}$ are in fm.}
\begin{center}
\begin{tabular}{cccccccccc}
\hline \hline
\mbox{} &  $\sqrt{\langle r^2 \rangle_B}$ & $\sqrt{\langle r^2 \rangle_E}$ & $\Delta_M$
& $M_{\rm sol}$ & $M_{\rm sol}^{O(p^2)}$ &  $M_{\rm sol}^{O(p^4)}$
& $M_{\rm sol}^{\rm anom}$ &  $M_{\rm sol}^{\rm dilaton}$ \\ \hline
HLS & 0.43 & 0.59 & 522.8 & 1188.8 & 878.4 & $-125.1$ & 435.4 & 0 \\
dHLS-I & 0.43 & 0.60 & 555.1 & 1138.0 & 746.2 & $-114.9$ & 458.0 & 48.8  \\
dHLS-II & 0.41 & 0.58 & 636.0 & 1099.1 & 696.0 & $-117.1$ & 431.4 & 89.0  \\
\hline \hline
\end{tabular}
\end{center}
\end{table*}

By solving the coupled equations of motion, one can calculate the properties of 
a single skyrmion.
Shown in Table~\ref{table:skyrmiondhls1} are the skyrmion properties obtained in 
dHLS-I and dHLS-II.
We present the results with the dilaton parameters given in Eq.~(\ref{eq:chi_param}).
For comparison, we show the results in HLS that has no dilaton field.
By varying the dilaton mass we could also confirm that, the heavier the dilaton mass,
the closer the results of dHLS-I and dHLS-II come to those of HLS, as anticipated.

The inclusion of the dilaton does indeed reduce the soliton mass, although not as 
much as needed.
The mass reduction is found to be $\sim 50$~MeV and $\sim 90$~MeV, respectively,
for dHLS-I and dHLS-II.%
\footnote{In the minimal model, this reduction is about 50~MeV~\cite{MRY99}.}
Given that the dilaton mass term itself increases the soliton mass by about the same
amount in magnitude, we see that the attraction due to the dilaton coupling to other
fields is twice the mass reduction, which is substantial.
Here, the main contribution comes from the factor $(\chi/f_\chi)^2$ in the Lagrangian
$\mathcal{L}_{(2)}$, which is less than 1 in the central region of the skyrmion.
On the other hand, in dHLS-I, the $\omega$ mass is reduced effectively by the same 
factor and provides more repulsion to the solution, as can be checked by the increase 
of the soliton mass from the hWZ terms.
As for the $N$-$\Delta$ mass difference denoted by $\Delta_M$ in
Table~\ref{table:skyrmiondhls1}, the dilaton causes its increase.
It is not desirable but can be understood from the fact that $\Delta_M$ is inversely
proportional to the moment of inertia with respect to the isospin rotation.
With the dilaton field, the factor $(\chi/f_\chi)^2$ in $\mathcal{L}_{(2)}$ causes 
smaller moment of inertia that leads to a larger $\Delta_M$.
On the other hand, the skyrmion size is almost unaffected by the presence of
the dilaton.
The rms radius of the soliton evaluated by weighting the baryon number density,
$\sqrt{\langle r^2\rangle_B}$, is almost unchanged by the incorporation of the dilaton,
and the energy density weighted rms radius $\sqrt{\langle r^2\rangle_E}$ shows only
a slight change.
The breakdown of the soliton mass in each model is shown in 
Table~\ref{table:skyrmiondhls1}.


\begin{figure*}[t] \centering
\includegraphics[width=0.6\textwidth]{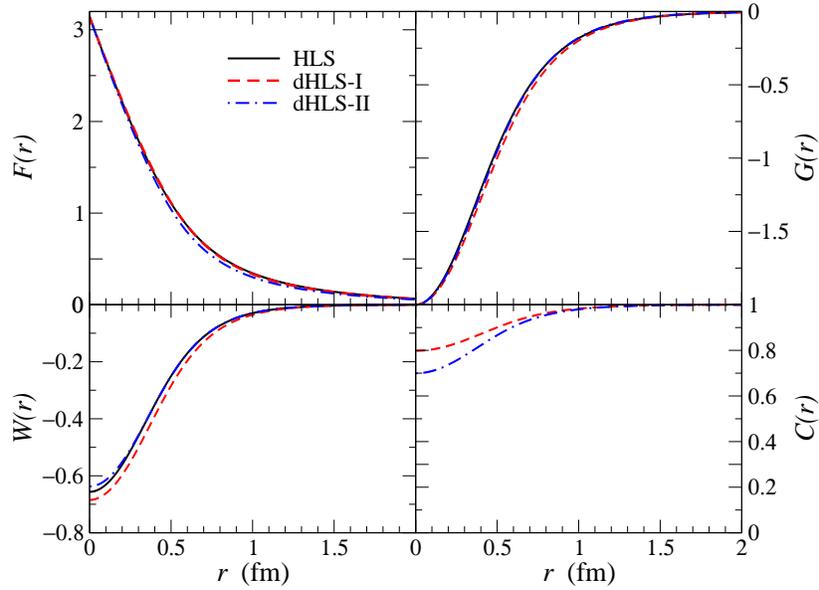}
\caption[]{Profiles of the wave functions in HLS (solid lines),
dHLS-I (dashed lines), and dHLS-II (dot-dashed lines).
In HLS, $C(r) = 1$ and it is not drawn here.}
\label{fig:ProfiledHLS}
\end{figure*}

The profiles of the wave functions of dHLS-I and dHLS-II are shown by dashed and 
dot-dashed lines, respectively, in Fig.~\ref{fig:ProfiledHLS}.
For comparison, the wave functions of HLS are also given by solid lines, for which 
$C(r)=1$ and hence is not drawn. 
One can find that the wave functions $F(r)$, $G(r)$, and $W(r)$ are almost unaffected 
by the presence of the dilaton field.
This explains that the skyrmion size is almost unaffected by the dilaton field, as 
is verified in Table~\ref{table:skyrmiondhls1}.
However, the changes in $W(r)$ show opposite behaviors in dHLS-I and dHLS-II.
We can understand such a change in $W(r)$ by the fact that the $\omega$ mass scales by the
factor $(\chi/f_\chi)$ in dHLS-I, becoming effectively lighter in the central region,
while it is not scaled in dHLS-II.
The wave function of the dilaton field $C(r)$ illustrates that the scale symmetry
--- and consequently the chiral symmetry --- is partially restored in the central 
region of the skyrmion, which is consistent with the chiral bag 
picture~\cite{BR79,BRV79}.


\section{Dense skyrmion matter in \lowercase{d}HLS-I and \lowercase{d}HLS-II}
\label{sec:matter}

\subsection{FCC skyrmion crystal}

Although the effect of the dilaton on a single skyrmion is relatively minor, 
the dilaton plays a far more important role in dense matter made of such skyrmions.
The dense skyrmion matter can be constructed by putting the skyrmions onto the 
FCC crystal sites following the procedure outlined, for example, in 
Refs.~\cite{MHLOPR13,LPMRV03,PRV03}.
To get the lowest energy configuration, the skyrmion at each lattice site should 
be arranged in such a way that the nearest skyrmions have the maximum attraction, 
for which the skyrmions at the closest sites should be relatively rotated in the 
isospin space by an angle $\pi$ about the axis perpendicular to the line joining them.
This requires that $\pi({\bm r})$, $\rho_\mu({\bm r})$, $\omega_\mu({\bm r})$, and
$\chi({\bm r})$ for the FCC crystal configuration%
\footnote{In this paper, we deal with the skyrmion crystal only at the leading order 
in $N_c$.}
should obey the periodic but distorted boundary conditions associated with the 
required symmetries with respect to the translation, reflection, fourfold rotations, 
and so on. (See Ref.~\cite{PRV03} for details.)

The classical solutions for $\pi$, $\rho$, $\omega$, and $\chi$ mesons satisfying 
the symmetries and carrying  a specified baryon number per box can be obtained by 
applying the Fourier expansion method developed in Ref.~\cite{KS88-KS89a} for the 
original Skyrme model, then generalized in Ref.~\cite{PRV03} for the model with 
vector mesons and dilaton.
For $\pi$, $\rho$, and $\omega$ fields, we use the convention of 
HLS($\pi, \rho, \omega$) described in Ref.~\cite{MHLOPR13}. 
The isoscalar dilaton field $\chi$ could be expanded as
\begin{eqnarray}
\frac{\chi({\bm r})}{f_\chi}  = \sum_{abc} 
\beta^\chi_{abc} \cos\left( \frac{\pi ax}{L} \right)
\cos\left( \frac{\pi by}{L} \right) \cos\left( \frac{\pi cz}{L} \right)
\label{eq:fourierC}
\end{eqnarray}
with the expansion coefficients $\beta_{abc}^\chi$ with the same integer set 
$(a,b,c)$ as that of the $\omega$ meson. 
Here, $L$ is the half length of the edge of a single FCC box containing four skyrmions.
The normal nuclear matter density $n_0^{} = 0.17/\mbox{fm}^3$ corresponds to 
the crystal size $L \sim 1.43$~fm.

The minimum energy configuration can be found numerically by taking the expansion 
coefficients as the variational variables.
However, the $\omega$ meson field needs a special treatment as in Ref.~\cite{PRV03}.
Since the $\omega$ meson provides a repulsive interaction and gives a positive definite
contribution to the energy, a straightforward variational process always ends up 
with the trivial results $\omega_0^{} = 0$.
It is nonetheless the correct solution to the equation of motion for the $\omega$ 
with the nonvanishing source term; viz.,
\begin{eqnarray}
\left[ - \partial_i \partial_i + C^2({\bm r}) m_\omega^2 \right] \omega_0({\bm r})
&=& S^\omega({\bm r}), 
\label{eq:eomwdhls}
\end{eqnarray}
where
\begin{equation}
C({\bm r}) = \left\{
\begin{array}{ll}
\chi({\bm r})/f_\chi & \quad \mbox{for dHLS-I}, \\
1 & \quad \mbox{for dHLS-II}.
\end{array}
\right.
\end{equation}
Both in dHLS-I and dHLS-II, the source term, $S^\omega$ in Eq.~\eqref{eq:eomwdhls}, 
comes from the hWZ terms,
\begin{eqnarray}
S^\omega &=&  - \frac{g_\omega^{} N_c}{32\pi^2} \varepsilon_{ijk}^{}
\Bigl[ (c_1^{} + c_2^{}) \, \tilde{\bm{\alpha}}_{\perp i }^{} \cdot
\left( \tilde{\bm{\alpha}}_{\parallel j}^{} \times \tilde{\bm\alpha}_{\parallel k}^{} \right)
\nonumber\\
&&  \quad  \mbox{}
+ (c_1^{} - c_2^{}) \, \tilde{\bm\alpha}_{\perp i }^{} \cdot
\left( \tilde{\bm\alpha}_{\perp j }^{} \times \tilde{\bm\alpha}_{\perp k }^{} \right)
\nonumber\\
&& \quad  \mbox{}
- 2 c_3^{} \left\{ \bm{V}_{ij} \cdot \tilde{\bm{\alpha}}_{\perp k }^{} -
\varepsilon_{ijk} \, \partial_i (
\tilde{\bm\alpha}_{\parallel j}^{} \cdot \tilde{\bm\alpha}_{\perp k}^{} ) \right\} \Bigr] .
 \label{eq:sourcew}
\end{eqnarray}
Thanks to the symmetries of the fields, the source term can be expanded in the 
Fourier series with the same set of the integers $(a,b,c)$ as those of 
$\omega({\bm r})$ and can be written as
\begin{equation}
S^\omega({\bm r}) =
\sum_{abc} \gamma_{abc}^{} \cos\left(\frac{\pi ax}{L}\right)
\cos\left(\frac{\pi by}{L}\right) \cos\left(\frac{\pi cz}{L}\right).
\end{equation}
Then, the equation of motion for the $\omega$ can be reduced to a linear matrix 
equation for $\beta^\omega_{abc}$ as
\begin{equation}
\sum_{a' b' c'} D_{abc, a' b' c'}^{} \beta^\omega_{a' b' c'} = \gamma_{abc}^{} ,
\end{equation}
where the matrix elements $D^{-\partial_i^2}_{abc, a' b' c'}$  from
the Laplacian $\-\partial_i^2$ and $D^{m_\omega^2}_{abc, a' b' c'}$
from the $\omega$ mass term in dHLS-II form diagonal matrices,
\begin{eqnarray}
D^{-\partial_i^2}_{abc, a' b' c'} &=& (a^2+b^2+c^2)
\left(\frac{\pi}{L}\right)^2 \delta_{aa'} \delta_{bb'} \delta_{cc'} ,
\nonumber \\
D^{m_\omega^2}_{abc, a' b' c'} &=&
m_\omega^2 \delta_{aa'} \delta_{bb'} \delta_{cc'}.
\end{eqnarray}
The matrix $D^{C^2 m_\omega^2}$ from the space-dependent $\omega$ mass term
with $C^2({\bm r})$ in dHLS-I is nondiagonal and its element has the form of
\begin{equation}
D^{C^2 m_\omega^2}_{abc,a' b' c'} = m_\omega^2 \sum_{a'',b'',c''}
\beta^{C^2}_{a''b''c''} f_{a' a'' a} f_{b' b'' b} f_{c' c'' c}
\end{equation}
with the Fourier expansion coefficients $\beta^{C^2}_{abc}$ for $C^2({\bm r})$ and
\begin{equation}
f_{a' a'' a} = \left\{
\begin{array}{ll}
\delta_{a' a} & \mbox{ if } a''=0, \\
\delta_{a'' a} & \mbox{ if } a'=0, \\
\frac12 \delta_{a, a' \pm a''} & \mbox{ if } a' a'' \neq 0.
\end{array}
\right.
\end{equation}
Finally, the Fourier expansion coefficients $\beta^{\omega}_{abc}$ can be obtained by
multiplying the inverse matrix $D^{-1}$ to $\gamma_{abc}^{}$.

\begin{figure}[t] \centering
\includegraphics[width=0.95\columnwidth]{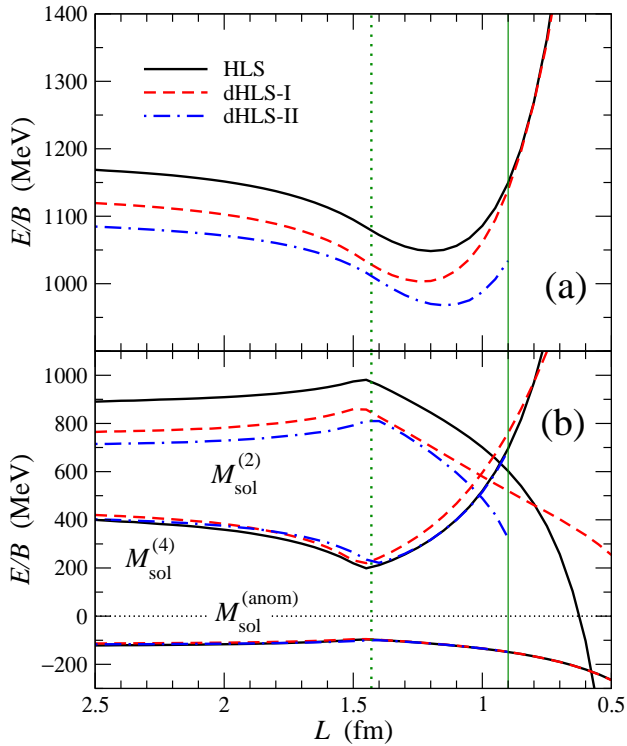}
\caption[]{(a) Energy per baryon $E/B$ as a function of crystal size $L$.
(b) Contributions from $\mathcal{L}_{(2)}$, $\mathcal{L}_{(4)}$, and 
$\mathcal{L}_{\rm anom}$ to $E/B$.
The results of dHLS-I and dHLS-II are given by the dashed and dot-dashed lines, 
respectively, while those of HLS are shown by solid lines for comparison.
The vertical dotted line shows the position of normal nuclear density, i.e., 
$L = 1.43$~fm, and the vertical solid line shows that of the critical density 
$n_c^{}$ that corresponds to $L \simeq 0.9$~fm.}
\label{fig:energydHLS}
\end{figure}

In Fig.~\ref{fig:energydHLS}, we present the obtained energy per baryon $(E/B)$ 
as a function of the crystal size $L$.
The contributions from $\mathcal{L}_{(2)}$, $\mathcal{L}_{(4)}$, and
$\mathcal{L}_{\rm anom}$ to $E/B$ are also presented.
The results from dHLS-I and dHLS-II are shown by dashed and dot-dashed lines,
respectively.
For comparison, HLS results are also shown by solid lines.
Besides the overall reduction in $E/B$ due to the changes in the single skyrmion 
mass as discussed in the previous section, we can see that, in the case of dHLS-I,
$n_{\rm min}^{}$ where $E/B$ has the minimum value is slightly moved to a lower 
value than that of HLS.
In the case of dHLS-II, there is no noticeable change in $n_{\rm min}^{}$ but $E/B$ 
drops suddenly at a density denoted by $n_c^{} \simeq 4\, n_0^{}$ whose position 
is given by the vertical solid lines in Fig.~\ref{fig:energydHLS}, i.e., at 
$L \simeq 0.9$~fm.
As we will see later, at this density, not only does the overall average of the dilaton field
vanish but also $\chi({\bm r}) = 0$ in the whole space.
In Fig.~\ref{fig:energydHLS}(b), we can see explicitly the effect of the dilaton 
on the $\mathcal{L}_{(2)}$, $\mathcal{L}_{(4)}$, and $\mathcal{L}_{\rm anom}$ 
contributions to $E/B$.
One can see that they have similar density dependence.
It shows clearly that, as in the case of a single skyrmion, the dilaton field 
mainly affects the contributions from $\mathcal{L}_{(2)}$.
Again, this reflects that the dilaton couples only to the terms of $\mathcal{L}_{(2)}$.


\begin{figure}[t] \centering
\includegraphics[width=0.95\columnwidth]{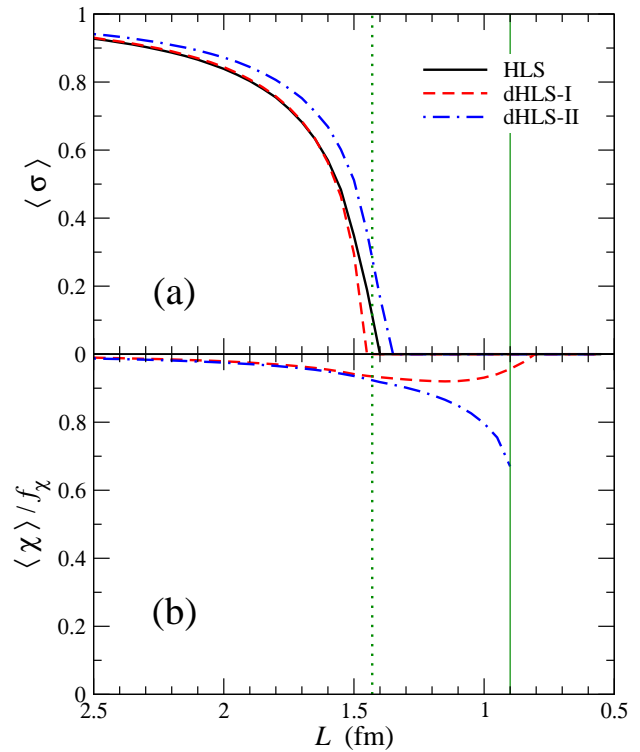}
\caption[]{(a) The expectation value $\langle \sigma \rangle$ and
(b) $\langle \chi \rangle/f_\chi$ as functions of crystal size $L$.
The notations are the same as in Fig.~\ref{fig:energydHLS}.}
\label{fig:vevdHLS}
\end{figure}

Figure~\ref{fig:vevdHLS} shows the space averaged quantities, 
$\langle \sigma \rangle$ and $\langle \chi \rangle$, as functions of the crystal 
size $L$.
Here, $\sigma$ is defined by $U=\xi^2 = \sigma + i {\bm \tau} \cdot {\bm \phi}_\pi$ 
and the space average $\langle A \rangle$ of quantity $A(r)$ is
\begin{equation}
\langle A \rangle = \frac{1}{V_{\rm box}} \int_{\rm box} d^3 {\bm r} \, A({\bm r}),
\end{equation}
where the integral is over a single FCC box with the volume $V_{\rm box} = 8L^3$.
The vanishing of $\langle \sigma \rangle$ signals the skyrmion--half-skyrmion 
phase transition.
In Fig.~\ref{fig:vevdHLS}(a) it is shown that the density of the half-skyrmion 
phase transition $n_{1/2}^{}$ is changed with the inclusion of the dilaton.
Interestingly, they are opposite in dHLS-I and dHLS-II; in dHLS-I, $n_{1/2}^{}$ 
becomes slightly lower than that of HLS, while it becomes slightly higher in dHLS-II.

As can be seen in Fig.~\ref{fig:vevdHLS}(b), the dependence of $\langle \chi \rangle$ 
on the crystal size at high density is completely different in dHLS-I and dHLS-II.
In dHLS-I, as density increases, $\langle \chi \rangle$ decreases until the density
approaches to about $n_{\rm min}^{}$, but after that it begins to increase.
Such an increase in $\langle \chi\rangle$ has been reported in the previous work
of Ref.~\cite{PRV03}.
Once we accept  that the scale symmetry is locked to the chiral symmetry, we expect
$\langle \chi \rangle$ to decrease, not increase, as density increases.
The main reason for this behavior comes from the $\chi^2$ term that we have 
introduced into the second term in $\mathcal{L}_{(2)}$, which makes the $\omega$ 
mass scale with $\chi$. 
The contribution of the hWZ term to $E/B$ can be approximately expressed as
\begin{eqnarray}
&& (E/B)_{\rm anom} = 
\nonumber \\ &&
\frac{1}{4} \int_{\rm box}d^3 r\int d^3 r^\prime 
S^\omega({\bm r})\frac{\exp(-m_\omega^{\ast}|{\bm r} - {\bm r}'|)}
{4 \pi|{\bm r} - {\bm r}'|}S^\omega({\bm r}'),
\label{eq:perehwzdhlsi}
\end{eqnarray}
where $m_\omega^*$ is the ``effective" $\omega$ mass.
Note that the integration over ${\bm r}$ is restricted in the single FCC box but 
that over ${\bm r}'$ is over all the space.
In order to make $(E/B)_{\rm anom}$ finite, the screening through a nonvanishing
$\omega$ mass is unavoidable.

The model based on dHLS-II, where $\mbox{[U(2)]}_{\rm HLS}$ symmetry is broken 
down to $[\mbox{SU(2)} \times \mbox{U(1)}]_{\rm HLS}$, avoids the above-mentioned 
difficulty.
We see in Fig.~\ref{fig:vevdHLS}(b) that $\langle \chi \rangle$ smoothly decreases 
to, and beyond, $n_{1/2}^{}$ and drops rapidly to zero when a higher density 
$n_c^{}$ is reached.%
\footnote{As discussed in Ref.~\cite{PRV08}, an alternative way to get finite 
$(E/B)_{\rm anom}$ is to introduce a scale-dependent $g_\omega^{}$ to weaken the 
source itself.
In Ref.\cite{PRV08}, this was realized by multiplying the 
$S^\omega$ term by the factor $\chi^3$ so that the decrease in the effective $\omega$ mass is 
accompanied by the decrease in the effective source. 
However, it is found that the weakening of the $\omega$ coupling upsets the 
stability of the single skyrmion for a light dilaton that is needed for nuclear 
phenomenology.
As a variation along this direction, one can endow an explicit density dependence 
in $g_\omega^{}$.
In this case,  there is no problem with the single skyrmion properties.
This will be described in the discussion section.}


\subsection{In-medium properties of mesons}

Once the skyrmion crystal is constructed, one can use it to study the in-medium 
properties of mesons as proposed in Refs.~\cite{LPMRV03,LPRV03a}.
Taking the skyrmion crystal solution as background classical fields, we can 
interpret the fluctuating fields on top of it as the corresponding mesons in 
dense baryonic matter.
For this purpose, we denote the minimum energy solutions as $\xi_{(0)}({\bm r})$,
$\rho^{a(0)}_{\mu}$, $\omega_{\mu}^{(0)}$, and $\chi^{(0)}$,
and then introduce the fluctuating fields as
\begin{eqnarray}
\xi_{L,R}^{} &=&\tilde{\xi}_{L,R}^{} \xi_{(0)L,R}^{}  , 
\nonumber\\
V_\mu^{(\rho),a} &=& \frac{1}{2} g_\rho^{} \rho_\mu^{a(0)}
+ \frac{1}{2} g_\rho^* \tilde{\rho}_\mu^a, 
\nonumber\\
V_\mu^{(\omega)} &=& \frac{1}{2} g_\omega^{} \omega_\mu^{(0)}
+ \frac{1}{2} g_\omega^* \tilde{\omega}_\mu^{},
\nonumber\\
\chi &=& \chi^{(0)} + \tilde{\chi} , 
\label{eq:fieldfluct}
\end{eqnarray}
where $\tilde{\xi}^\dagger_L = \tilde{\xi}_R^{} = \tilde{\xi} 
= \exp (i \tau_a \tilde{\pi}_a/2f_\pi)$, $\tilde{\rho}^a_\mu$, $\tilde{\omega}_\mu$, 
and $\tilde{\chi}$ stand for the corresponding fluctuating fields. 
In Eq.~\eqref{eq:fieldfluct}, $g_\rho^*$ and $g_\omega^*$ are the medium modified 
HLS gauge couplings of the $\rho$ and $\omega$ mesons, respectively. 
It is worth noting that the decomposition given in Eq.~\eqref{eq:fieldfluct} can 
easily keep the HLS of the matter in terms of the expansion of the quantum fluctuations 
by imposing that the fluctuations transform homogeneously under the HLS, but the 
matter fields transform the same as their corresponding original quantities in HLS. 
By substituting the fields in Eq.~\eqref{eq:fieldfluct} into the dHLS Lagrangian,
one can obtain the medium modified one.

To define the pion decay constant in the skyrmion matter, we consider the 
axial-vector current correlator
\begin{eqnarray}
i G_{\mu\nu}^{ab}(p) & = & i \int d^4 x \, e^{ip\cdot x}
\left\langle 0 \mid TJ_{5\mu}^a(x)J_{5\nu}^b(0) \mid 0\right\rangle .
\label{eq:defaacorr}
\end{eqnarray}
This correlator can be evaluated from the medium modified Lagrangian by introducing 
the corresponding external source by gauging the chiral symmetry, i.e., substituting 
the covariant derivative defined in Eq.~\eqref{eq:covDhls} with
\begin{eqnarray}
D_\mu \xi_{L}^{} & = & (\partial_\mu - i V_\mu) \xi_{L}^{} 
+ i \xi_{L,R}^{} \mathcal{L}_\mu , 
\nonumber\\
D_\mu \xi_{R}^{} & = & (\partial_\mu - i V_\mu) \xi_{R}^{} 
+ i \xi_{L,R}^{} \mathcal{R}_\mu ,
\end{eqnarray}
where ${\cal L}_\mu$ and ${\cal R}_\mu$ are introduced as the gauge fields of the 
chiral symmetry. 
The external source of the axial-vector current $J_{\mu 5}$ is a combination 
$(\mathcal{R}_\mu - \mathcal{L}_\mu)/2$.


\begin{figure}[t] \centering
\includegraphics[width=1.\columnwidth]{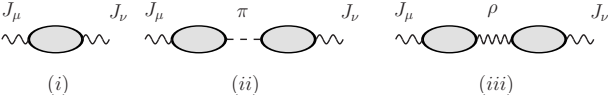}
\caption[]{Three types of contributions to the correlator of Eq.~\eqref{eq:defaacorr}: 
(i) the contact diagram, 
(ii) the pion exchange diagram, and 
(iii) the $\rho$ exchange diagram. 
Shaded blobs stand for the skyrmion matter interaction vertices.}
\label{fig:Correlator}
\end{figure}

In the present calculation, we do not consider the contributions from the loop diagrams 
of the fluctuation fields to the correlator of Eq.~\eqref{eq:defaacorr}. 
Therefore, as illustrated in Fig.~\ref{fig:Correlator}, there are three types of contributions: 
(i) the contact diagram, 
(ii) the pion exchange diagram, and (iii) the $\rho$ exchange diagram. 
In the present evaluation of the correlator, we only consider the matter effect 
from $\xi_{(0)L,R}$ and $\chi^{(0)}$ but leave a complete calculation, including 
derivative on them, to our future publication. 
In such an approximation the three types of contributions are expressed as
\begin{eqnarray}
&\mbox{(i)} : & i f_\pi^2 g_{\mu\nu}^{} \delta^{ab}
\left\langle \frac{\chi_{(0)}^2}{f_\chi^2} \left\{1 + 
\frac{1-a}{2} \left[\left(1-\frac{2}{3}\bm{\phi}_\pi^2\right) - 1\right]\right\} 
\right\rangle ,
\nonumber\\
&\mbox{(ii)} : &
- i f_\pi^2 \frac{p_\mu^{} p_\nu^{}}{p^2} \delta^{ab}
\left\langle \frac{\chi_{(0)}^2}{f_\chi^2}\left[\left(1-\frac{2}{3}\bm{\phi}_\pi^2
\right) - 1\right] \right\rangle ,
\nonumber\\
&\mbox{(iii)} : &
 i \delta^{ab}\left\langle \frac{\chi_{(0)}^2}{f_\chi^2}
\frac{a^2g^2f_\pi^4}{p^2 - \frac{\chi_{(0)}^2}{f_\chi^2} m_\rho^2} 
\left(g_{\mu\nu} - \frac{p_\mu p_\nu}{\frac{\chi_{(0)}^2}{f_\chi^2} m_\rho^2}\right) 
\right. 
\nonumber\\
&&\left. \qquad \mbox{} \times \frac{\chi_{(0)}^2}{f_\chi^2} 
\left[\left(1-\frac{2}{3}\bm{\phi}_\pi^2\right) - 1\right] \right\rangle .
\end{eqnarray}
Summing over the above three types of contributions, one concludes that, to the 
leading order of the $p^2/(\chi_{(0)}^2m_\rho^2/f_\chi^2)$ expansion, the 
axial-vector current correlator \eqref{eq:defaacorr} is gauge invariant and 
therefore can be decomposed into the longitudinal and transverse parts as
\begin{eqnarray}
G_{\mu\nu}^{ab}(p) &=& \delta^{ab} \left[ P_{T\mu\nu} G_{T}(p) 
+ P_{L\mu\nu} G_{L}(p) \right],
\end{eqnarray}
where the polarization tensors $P_{L,T}$ are defined as
\begin{eqnarray}
P_{T\mu\nu}^{} &=& 
g_{\mu i}^{} \left( \delta_{ij} - \frac{p_i^{} p_j^{}}{|\bm{p}|^2} \right) g_{j\nu} ,
\nonumber\\
P_{L\mu\nu}^{} &=& - \left( g_{\mu\nu} - \frac{p_\mu^{} p_\nu^{}}{p^2}
\right) - P_{T\mu\nu}^{}.
\end{eqnarray}
We next define the medium modified pion decay constant through the longitudinal 
component in the low energy limit
\begin{eqnarray}
f_\pi^{\ast 2} & \equiv & {} - \lim_{p_0 \to 0} G_{L}(p_0,\bm{p}=0) 
\nonumber\\
& = & f_\pi^2 \left\langle \frac{\chi_{(0)}^2}{f_\chi^2}\left [1 - \frac{2}{3}
\left( 1-  \sigma^2_{(0)} \right)\right]\right\rangle ,
\label{eq:mediumfpi}
\end{eqnarray}
where the intrinsic density dependence is brought in by the minimal energy solution
$(\chi_{(0)}/f_\chi)^2$ and $\sigma_{(0)}^2$, and the relation 
$\sigma_{(0)}^2 + {\bm \phi}_\pi^2 = 1$ has been used. 
Note that, because of the rho meson exchange effect, the medium modified $f_\pi$ 
is independent of the HLS parameter $a$.

In our present approach, since only the $O(p^2)$ terms of HLS are considered, we 
should have
\begin{eqnarray}
g_\rho^{\ast} = g_\rho^{} 
\label{eq:grhostargrho}
\end{eqnarray}
for the normalization of the $\rho$ meson field in medium. 
Thus, because of the dilaton compensator,  the $\rho$ meson mass is modified to be
\begin{eqnarray}
\nonumber\\
m_\rho^{\ast 2} &=& \left\langle \frac{\chi_{(0)}^2}{f_\chi^2} \right\rangle m_\rho^2  .
\label{eq:scalmesondhlsi}
\end{eqnarray}
As for the $\omega$ meson mass, dHLS-I and dHLS-II lead to different results:
\begin{equation}
m_\omega^{\ast 2} = \left\{
\begin{array}{ll}
\displaystyle \left\langle \frac{\chi_{(0)}^2}{f_\chi^2} \right\rangle m_\omega^2 & 
\mbox{for dHLS-I}, \\
\quad m_\omega^2  & \mbox{for dHLS-II}.
\end{array}
\right.
\label{eq:scalingdHLSIIomega}
\end{equation}

From Eqs.~\eqref{eq:scalmesondhlsi} and \eqref{eq:scalingdHLSIIomega} one may 
think that the BR scaling for the $\rho$ and $\omega$ meson masses is reproduced
in dHLS-I~\cite{BR91}. 
However, as will be shown below, since $\langle \chi \rangle$ increases with 
increasing density above $n_{1/2}^{}$, the masses increase so they are not 
consistent with the BR scaling for $n \geq n_{1/2}^{}$.
Also the pion decay constant scales differently from that of the vector meson masses.


\begin{figure}[t] \centering
\includegraphics[width=0.95\columnwidth]{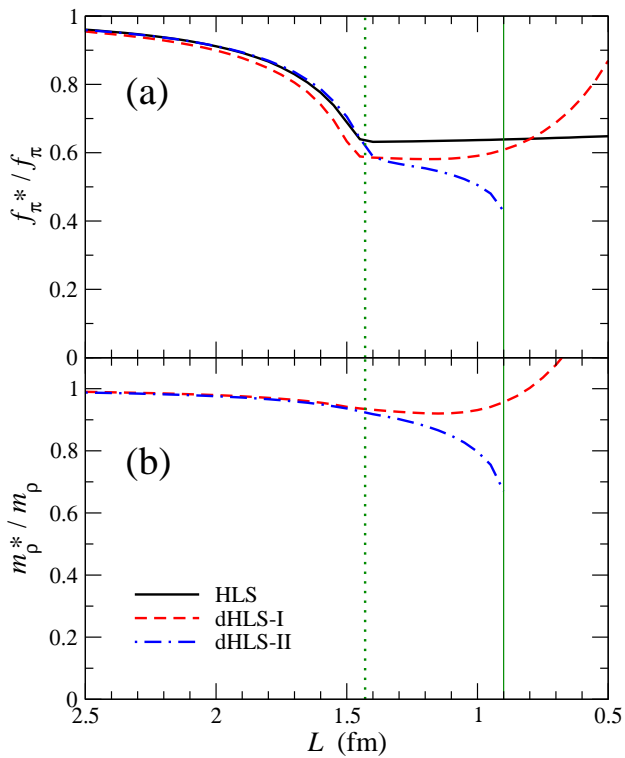}
\caption[]{The dependence of (a) $f_\pi^*/f_\pi$ and (b) $m_\rho^*/m_\rho^{}$ 
on the crystal size $L$.
The notations are the same as in Fig.~\ref{fig:energydHLS}.}
\label{fig:fpimrhodHLS}
\end{figure}

Plotted in Fig.~\ref{fig:fpimrhodHLS} are $f_\pi^{\ast}/f_\pi$ and $m_\rho^{\ast}/m_\rho^{}$
that show their dependence on the crystal size.
Since $m_\omega^{\ast}/m_\omega^{}$ is equal to $m_\rho^{\ast}/m_\rho^{}$ in dHLS-I 
or it does not scale in dHLS-II, we do not show it here.
As for $f_\pi^\ast/f_\pi$, for density up to $\sim n_{1/2}^{}$, the scaling behavior 
is mainly governed by $\sigma_{(0)}$, so that both in dHLS-I and in dHLS-II it 
decreases smoothly down to $f_\pi^\ast /f_\pi \sim 2/3$.

For density larger than $n_{1/2}^{}$, the scaling behavior is governed by
$\chi_{(0)}^2/f_\chi^2$, for which dHLS-I and dHLS-II yield different results.
In both cases, $f_\pi^{\ast} /f_\pi$ stays $\sim 2/3$ for a while.
Then, after $n_{\rm min}^{}$ it starts to increase in dHLS-I, but it goes down at 
higher density and then drops to zero at $n_c^{}$ in dHLS-II.
The in-medium $\rho$ meson mass $m_\rho^{\ast}$ scales only with
$\sqrt{\langle \chi_{(0)}^2/f_\chi^2\rangle}$ that is similar to 
$\langle \chi_{(0)}/f_\chi \rangle$, so it decreases monotonically until
$\sim n_{\rm min}^{}$, after which it starts to increase in dHLS-I, although it 
will ultimately drop to zero at $n_c^{}$ in dHLS-II.

The situation in dHLS-II is much simpler and appealing over all the range of density.
The quantities tied to chiral symmetry, $f_\pi^{\ast}/f_\pi$ and 
$m_\rho^{\ast}/m_\rho^{}$, do vanish at $n_c^{}$, showing that chiral symmetry is 
restored and the vector manifestation is realized.
It is intriguing that the VM is realized at the expense of breaking the
$\mbox{[U(2)]}_{\rm HLS}$ symmetry in medium and letting the $\omega$ meson mass
remain unscaled.

\subsection{The half-skyrmion phase}

The half-skyrmion phase exhibits some unusual properties of hadrons. 
This may be indicative of a non-Fermi liquid structure mentioned above~\cite{LR13c}. 
As one can see in  Fig.~\ref{fig:fpimrhodHLS}, in the phase for $n \geq n_{1/2}^{}$,
$\langle \sigma \rangle = 0$ but $f_\pi^{\ast} \neq 0$.
At the edge of the half-skyrmion phase, say, at $n_c^{}$, the half-skyrmion phase 
presumably transits to the Wigner phase with $\langle \chi \rangle = 0$ and 
$f_\pi^{\ast} = 0$.
In the Wigner phase the chiral symmetry is restored.

A question that arises is how to formulate the vector manifestation starting from the
half-skyrmion phase.
In matching the HLS and QCD correlators in arriving at the VM~\cite{HY03a,HY00}, 
the condition $\langle \bar{q}q \rangle \to  0$ plays a key role.
Now in the half-skyrmion phase, $\langle \bar{q}q \rangle=0$ but the pion decay 
constant is not zero. 
So we see that the VM cannot be realized in the half-skyrmion phase.
Furthermore, since $g(m_\rho^{}) \neq 0$ because of $m_\rho^{2} = 
a(m_\rho^{}) g_\rho^2 (m_\rho^{}) f_\pi^{2}(m_\rho^{}) \neq 0$, the Georgi's 
``vector realization'' of chiral symmetry~\cite{Geo89,Geo90a} cannot be arrived at.

Now let us approach the Wigner phase from the half-skyrmion phase.
At density $n_c^{}$, the effective pion decay constant and the $\rho$ meson mass 
take the values
\begin{eqnarray}
m_\rho^{\ast}(n_c^{}) = 0, \quad 
f_\pi^{\ast} \bm{(} m_\rho^{\ast}(n_c^{}) \bm{)} = f_\pi^{\ast}(0) = 0 ,
\end{eqnarray}
which can be regarded as the conditions to realize VM in medium.
In this sense, we can say that the VM of Eq.~\eqref{eq:VM} \textit{could be 
realized} in the model of dHLS-II.

It has been discussed in the literature~\cite{Zarembo01,PRV08} that there is a 
possible pseudo--gap phase in QCD in which quarks condensate and acquire constituent 
mass, but chiral symmetry is not broken because the condensate phase is completely 
disordered.
This situation is very similar to what happens in the half-skyrmion phase where 
the space average of quark-antiquark condensate vanishes and the $\rho$ is still 
massive, indicating that the quark acquires a constituent mass.
Thus if the order parameter of QCD were interpreted as the space averaged 
quark-antiquark condensate as in Refs.~\cite{Zarembo01,PRV08}, the half-skyrmion 
phase can be taken as the pseudo--gap phase. 
It could also be ``quarkyonic.''%
\footnote{But we know that at least on the crystal lattice, the quark condensate is 
not a good order parameter for chiral symmetry.}
Note that, in the half-skyrmion phase, the quark condensate is locally nonzero.

\subsection{In-mdeium baryon properties}

\begin{figure}[t] \centering
\includegraphics[width=0.95\columnwidth]{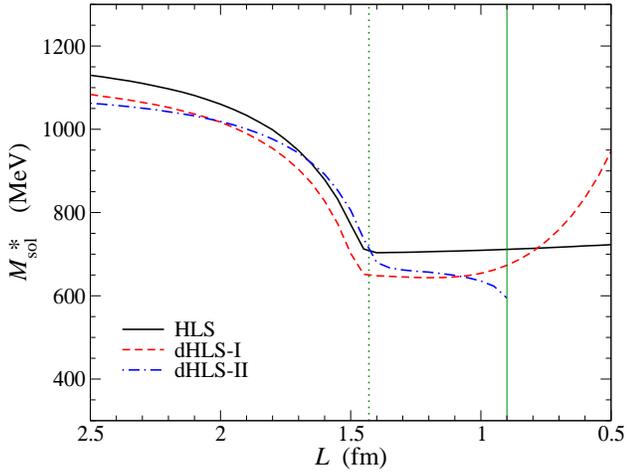}
\caption[]{In-medium modified skyrmion mass as a function of $L$.
The notations are the same as in Fig.~\ref{fig:energydHLS}.}
\label{fig:DenseSoliII}
\end{figure}

Substituting the inputs $f_\pi$, $m_\rho$, and $m_\omega$ with the corresponding 
medium modified $f_\pi^\ast$, $m_{\rho}^\ast$, and $m_\omega^\ast$ into the 
master formula~\eqref{eq:lecshls}, one can obtain the in-medium LECs of HLS.
Using these in-medium  LECs we can then calculate the intrinsic density-dependent 
nucleon mass.
We plot in Fig.~\ref{fig:DenseSoliII} the effective mass $M_{\rm sol}^\ast$ to see 
its density dependence.


From Fig.~\ref{fig:DenseSoliII}, we see that the soliton mass, which is identified 
as the nucleon mass in the large $N_c$ limit, can be parametrized as
\begin{eqnarray}
m_N^\ast & = & m_0^{} + \Delta(\langle \bar{q}q \rangle),
\end{eqnarray}
where $\Delta$ is the part of the nucleon mass arising from $\langle \bar{q}q \rangle$
that vanishes at $n = n_{1/2}^{}$ and $m_0^{}$ is the chiral-invariant mass, 
reminiscent of what figures in the parity doublet picture of baryons~\cite{DK89}.

We next make a parametrization of the scalings of $f_\pi^\ast$, $m_\rho^\ast$, 
and $m_N^\ast$ from our results shown in Figs.~\ref{fig:fpimrhodHLS} and 
\ref{fig:DenseSoliII}.
Such a parametrization would make some results of the present work easier to apply, 
for example, to the nuclear force and the EoS of nuclear matter~\cite{LR13c}. 
Here we only consider the results from dHLS-II.
Note that although $f_\pi^*$ and $m_N^*$ scale similarly, $m_\rho^{\ast}$ scales in a
different way.

We can roughly parametrize the medium-modified pion decay constant $f_\pi^\ast$ 
illustrated in Fig.~\ref{fig:fpimrhodHLS}(a) and the in-medium nucleon mass 
$m_N^\ast$ illustrated in Fig.~\ref{fig:DenseSoliII} as
\begin{eqnarray}
\frac{m_N^{\ast}}{m_N} \simeq \frac{f_\pi^{\ast}}{f_\pi} \simeq
\left \{\begin{array}{cl}
\displaystyle \frac{1}{1 + 0.6 (n/n_0^{})^2} \quad & \mbox{ for } n < n_{1/2}^{} ,\\
0.63  & \mbox{ for } n_{1/2} < n < n_c^{} ,\\
0 & \mbox{ for } n > n_c^{} .
\end{array}
\right.
\nonumber \\
\end{eqnarray}
For the in-medium vector meson mass $m_\rho^*$ in Fig.~\ref{fig:fpimrhodHLS}(b)
we parametrize it as
\begin{eqnarray}
\frac{m_\rho^{\ast}}{m_\rho} \simeq \left \{\begin{array}{cl}
\displaystyle \frac{1}{1 + 0.4 (n/n_c)^2} \quad & \mbox{ for }  n < n_{c}^{} , \\
0 & \mbox{ for } n > n_c^{} .
\end{array}
\right.
\end{eqnarray}


\section{Summary of the results}
\label{sec:summary}

The series of work done with multi-skyrmions obtained from a chiral Lagrangian with 
vector mesons incorporated as hidden gauge fields to simulate dense baryonic matter 
revealed a number of features that were not observed in chiral models without vector 
mesons.
The model used in the present work is based on the Lagrangian written up to $O(p^4)$
in the chiral expansion, including the pion, the $\rho$ meson, and the $\omega$ meson, 
and is parameter free thanks to master formulas derived from the 5D holographic QCD action
that arises from gauge-gravity duality of hQCD, as well as from dimensional 
deconstruction starting from $\mbox{SU(2)}_L\times \mbox{SU(2)}_R$ current algebra.
This Lagrangian is considered to be as close as one can hope to reach the large 
$N_c$ limit of QCD properly.

To repeat the most remarkable results:
\begin{enumerate}
\item  The isosinglet vector meson $\omega$ plays a crucial role in the structure 
of both the elementary nucleon and multinucleon systems.
Given that the $\omega$ meson is in the topological term encoding anomaly, it 
cannot be properly, if at all, captured in models that have no explicit $\omega$ 
degree of freedom such as the famous Skyrme model or chiral perturbation theory.

\item  The density $n_{1/2}^{}$ at which the skyrmion--half-skyrmion transition, 
a generic feature of all the skyrmion models on crystal, takes place is found to 
be not far from the equilibrium nuclear matter density $n_0 \sim 0.17$ fm$^{-3}$.
It is therefore testable experimentally, such as through the medium modified kaon 
mass~\cite{Park:2009mg} and nuclear tensor force~\cite{Lee:2010sw}
(for a recent review, see, e.g., Ref.~\cite{Rho:2014dea}). 
Without the $\rho$ and $\omega$ fields, the transition takes place at much too low 
a density to be compatible with what is accurately known in normal nuclear matter,
and without the hWZ term --- i.e., without the $\omega$ meson--- it comes much 
too high to be relevant to nature.

\item\label{p1}  In medium, the effective pion decay constant $f_\pi^\ast$ encoding 
the \textit{intrinsic} density dependence is found to drop smoothly, roughly in 
consistency with chiral perturbation theory, to the density $n_{1/2}^{}$, but stops 
dropping at $n_{1/2}^{}$ and remains constant $\sim (60\%-80\%)$ of the free-space 
value in the half-skyrmion phase.

\item\label{p2}  The in-medium nucleon mass $m_N^\ast$ tracks closely the in-medium 
pion decay constant $f_\pi^\ast$ ( multiplied by a scale-invariant factor 
proportional to $\sqrt{N_c}$ ) which indicates that the large $N_c$ dominance holds 
in medium as it does in free space and stays constant $\sim (60\%-80\%)$ of the 
free-space value in the half-skyrmion phase for $n\geq n_{1/2}^{}$.
Given that $\langle \bar{q}q \rangle^*=0$ in the half-skyrmion phase, the constant 
nucleon mass must be a chirally invariant term in the Lagrangian.
This resembles the chiral-invariant $m_0^{}$ term in the parity-doublet baryon model.
In our calculation, such a term is not explicitly present in the Lagrangian with 
which the skyrmion crystal is constructed.
Therefore it could very well be a symmetry ``emergent" from many-body correlations 
different from what Glozman interprets as an intrinsic property of QCD in Ref.~\cite{Gloz12}.
\end{enumerate}

The items \ref{p1} and \ref{p2} get support from independent analyses based on the
one-loop RG flow with baryon HLS (BHLS) and mean-field approximation with
dilaton-implemented BHLS~\cite{PLRS13}.
This suggests that the qualitative structure of the half-skyrmion phase is correct.
However, as pointed out in the present work, there is a tension with the VM and BR scaling.
This is because, since the pion decay constant and the $\rho$ meson mass stay 
constant and do not tend to the VM fixed point, one cannot make the matching of 
the HLS correlators to those of QCD crucial to describe chiral restoration at a 
certain high density $n_c^{}$.

In this paper, we remove the obstacle to the VM found in the HLS crystal by the 
dilaton field associated with the spontaneous breaking of conformal symmetry.
We found that in order for the dilaton to retain the VM feature or BR scaling, the
$\mbox{[U(2)]}_{\rm HLS}$ symmetry for the $\rho$ and $\omega$ which seems to 
hold in matter-free space has to be broken in medium.
With the symmetry breaking induced in the vector meson mass rather than in the gauge
coupling as was done in Ref.~\cite{PLRS13}, all four features mentioned above were
retained and, in addition, the tension with the VM present in the HLS model 
(without the dilaton) could be removed.

We should point out that there are some caveats to the ``good" results mentioned above.
(1) The energy of the system is minimized at a larger density than the known equilibrium
density $n_0^{}$ with a binding energy much larger than the empirical value.
This is not surprising since we have here a large $N_c$ theory.
In standard nuclear many-body theory anchored on effective field theory, a similar
overbinding and higher saturation density are obtained unless one introduces three-
and multibody forces (see, e.g., Ref.~\cite{KD11}).
Whether this ``higher-order" effect is encoded in the crystal calculation needs to 
be clarified.
(2) In all cases considered, the density $n_{1/2}^{}$ comes below the empirical value of
$n_0^{}$, with the HLS-II giving $n_{1/2}^{}$ close to $n_0^{}$.
From the phenomenology discussed in Ref.~\cite{DKLMR12}, the density $n_{1/2}^{}$ most
likely relevant to nature should be $\lesssim 2 n_0^{}$.
However, it seems that there will be no difficulty even if $n_{1/2}^{}$ comes close to but
above $n_0^{}$.
(3) The half-skyrmion structure is a classical picture, already present in the 
skyrmion description of mass number 4 (see Ref.~\cite{BMS10}) and will surely be 
modified by quantum effects.


\section{Further Discussions}
\label{sec:dis}

It is interesting to compare what we have found in this paper to what have been 
seen in other developments, which are closely related to each other.


\begin{figure}[t] \centering
\includegraphics[width=0.95\columnwidth]{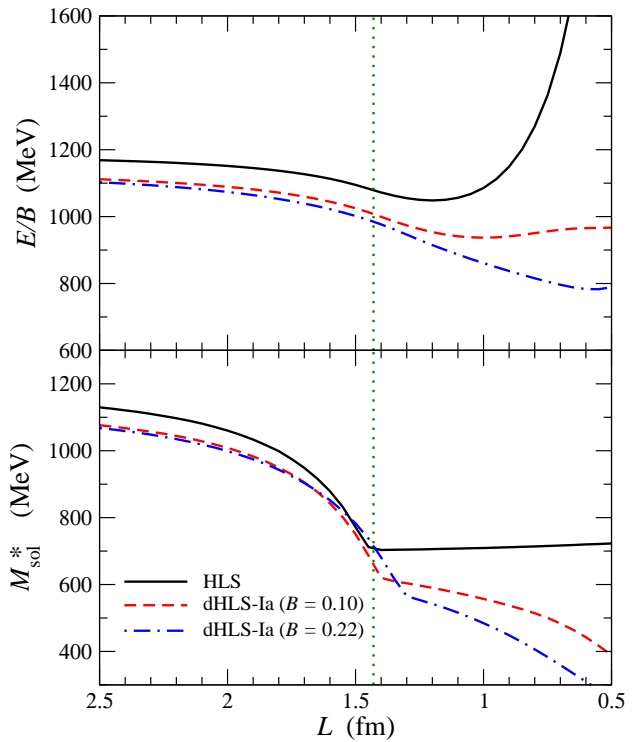}
\caption[]{ $E/B$ and $M_{\rm sol}^{*}$ in dHLS-Ia with $m_\chi = 720~$MeV.}
\label{fig:dhlsIII720}
\end{figure}

\begin{itemize}

\item[(i)]
Phenomenologically relevant is the application of the skyrmion--half-skyrmion
topology change to an effective nuclear field theory description of the EoS for 
compact stars~\cite{DKLMR12}.
Using the notion that topology change can be ``translated" into the parameter 
change of an effective Lagrangian, here in the form of the ``intrinsic density 
dependence" defined above, an effective nuclear Lagrangian was constructed by 
means of renormalization group equations and applied to calculating a high-order 
nuclear many-body problem.
The principal observation there was that in order to correctly describe nuclear 
matter and then extrapolate to higher density, it was essential that the effective 
nucleon mass drop smoothly to $\sim 0.8 \, m_N^{}$ up to density $\sim 2\, n_0^{}$ 
and then stay constant up to the density $\sim 5.5\, n_0^{}$ predicted to be 
present in the interior of a massive neutron star and that the $\omega NN$ coupling 
be more or less unscaled in the density regime involved.
This is roughly the feature obtained in dHLS-II.

\item[(ii)] 
Suppose we follow the development made in Ref.~\cite{PLRS13} using the 
mean-field approximation in dHLS-I, supplemented with explicit baryon degrees of 
freedom and $\mbox{[U(2)]}_{\rm HLS}$ symmetry breaking in the gauge coupling 
constants instead of in the vector meson masses.
In the dHLS-I($\pi,\rho,\omega$) crystal calculation, the corresponding procedure 
would be to replace $g_\omega^{}$ with a density-dependent one in the form of
\begin{eqnarray}
g_\omega^{} \to g_\omega^{} \frac{1}{1 + B (n/n_0^{})}.
\end{eqnarray}
where $n_0^{}$ is the normal nucleon density and $B$ is a parameter.
We refer to this model as dHLS-Ia and present the obtained results for the per 
skyrmion energy and $M_{\rm sol}^\ast$ in Fig.~\ref{fig:dhlsIII720} for two cases 
with $B = 0.10$ and $0.22$, respectively.
The dilaton mass is taken to be $m_\chi^{} = 720$~MeV, as before.
Our result shows that to arrive at the minimum of $E/B$, a smaller parameter $B$ 
is preferred.
It clearly shows that there is a density region after $n_{1/2}^{}$ in which the 
nucleon mass is nearly density independent. 
This agrees with the observation made in Ref.~\cite{PLRS13}.
Note that because of the density-dependent coupling $g_\omega^{}$, $n_{1/2}^{}$ 
is pushed to a higher density by the dilaton.

\item[(iii)] 
In the HLS calculation, we find that, in the half-skyrmion phase, the space 
average $\langle \sigma \rangle = 0$, but $\langle \sigma^2 \rangle \ne 0$ 
although it is a small density-dependent quantity. 
This observation might indicate that, although the space average quarkonia condensate 
vanishes in the half-skyrmion phase, the space average tetraquark condensate does not. 
Probably this is the first example which realizes the chiral symmetry breaking pattern 
$\mbox{SU}(N_f)_L \times \mbox{SU}(N_f)_R \to \mbox{SU}(N_f)_V \times \mbox{Z}(N_f)_A$ 
proposed in Ref.~\cite{Kogan:1998zc} \textit{emergent} in dense baryonic matter, whose 
features of the thermodynamic quantities and hadron mass spectra of this phase in the
two flavor case are explored in Ref.~\cite{Harada:2009nq}.    
If one assumes the Gell-Mann--Oakes--Renner relation holds in the half-skyrmion phase 
in the form ${f_\pi^*}^2 {m_\pi^*}^2 = D \langle\sigma^2\rangle^*$, with $D$ a 
density-independent constant, then since $\langle\sigma^2\rangle^*$ drops to a 
small value while $f_\pi^*$ remains more or less constant, one should expect 
$m_\pi^*$ should accordingly decrease fast in the half-skyrmion phase in the real 
world where the pion mass is nonzero. 
This would imply that the contribution from the pion exchange to the nuclear tensor 
forces will be enhanced for density $n\geq n_{1/2}$ and hence will increase the net 
tensor-force attraction, as is clear from the finding in Ref.~\cite{DKLMR12}. 
As was pointed out a long time ago by Pandharipande and Smith~\cite{pandha}, such an 
enhanced tensor force could lead to a p-wave $\pi^0$-condensed neutron solid at 
high density in compact stars.

\item[(iv)] 
A remarkable feature that has been uncovered in all models anchored on HLS is that
at a certain density above the normal nuclear matter density, the effective nucleon mass
$m_N^\ast$ stops being dependent on the chiral condensate, saturating at $\gtrsim 60\%$
of the free-space mass, and then stays constant until quark deconfinement.
We note that this is reminiscent of the chiral-invariant mass $m_0^{}$ posited in the
parity-doublet baryon model.
The mass $m_0^{}$ could be an intrinsic quantity of QCD proper in the sense suggested
in Ref.~\cite{Gloz12}.
However, the parity-doublet baryon model is an effective theory.
Now in our treatment, the chiral-invariant mass --- that breaks explicitly conformal invariance ---
is not put in {\it ab initio} in the Lagrangian.
Therefore, it is plausible that it rather reflects an emergent symmetry due to HLS skyrmion
interactions, and not an intrinsic one.
We note that since what we are looking at is a process of ``unbreaking symmetry" by
density, this indicates a subtle mechanism by which the nucleon mass could have been
generated.

\end{itemize}


\acknowledgments

We are grateful to APCTP for supporting the APCTP-WCU Focus program where this 
work was initiated.
This work was completed while three of us (Y.-L.M., B.-Y.P., and M.R.) were 
visiting the Rare Isotope Science Project (RISP) funded by the Ministry of 
Science, ICT, and Future Planning (MSIP) and National Research Foundation (NRF) 
of Korea.
We are grateful to Youngman Kim for making this visit feasible and Y.-L.M. 
acknowledges the support from the RISP.
\newblock
The work of Y.-L.M. and M.H. was supported in part by a Grant-in-Aid for Scientific 
Research on Innovative Areas (No. 2104) ``Quest on New Hadrons with Variety of 
Flavors'' from MEXT.
Y.-L.M. was supported in part by the National Science Foundation of China (NSFC) 
under Grant No.~10905060.
\newblock
The work of M.H. was partially supported by the Grant-in-Aid for Nagoya University 
Global COE Program ``Quest for Fundamental Principles in the Universe: From 
Particles to the Solar System and the Cosmos'' from MEXT, the JSPS Grant-in-Aid 
for Scientific Research (S) No. 22224003 and (c) No. 24540266.
\newblock
The work of H.K.L. and M.R. was partially supported by the WCU project of 
Korean Ministry of Education, Science and Technology (R33-2008-000-10087-0).
\newblock
Y.O. was supported by the Basic Science Research Program through the National
Research Foundation of Korea under Grant No. NRF-2013R1A1A2A10007294.
B.-Y.P. was supported by the research fund of Chungnam National University.


\end{document}